\pdfoutput=1
\documentclass[iop]{emulateapj}

\slugcomment{}

\shorttitle{Star formation activity toward the bubble N46}
\shortauthors{L.~K. Dewangan et al.}

\begin{document}

\title{Star formation activity in the neighbourhood of W-R 1503-160L star in the mid-infrared bubble N46}
\author{L.~K. Dewangan\altaffilmark{1}, T. Baug\altaffilmark{2}, D.~K. Ojha\altaffilmark{2}, P. Janardhan\altaffilmark{1}, J.~P. Ninan\altaffilmark{2}, A. Luna\altaffilmark{3}, and I. Zinchenko\altaffilmark{4}}
\email{lokeshd@prl.res.in}
\altaffiltext{1}{Physical Research Laboratory, Navrangpura, Ahmedabad - 380 009, India.}
\altaffiltext{2}{Department of Astronomy and Astrophysics, Tata Institute of Fundamental Research, Homi Bhabha Road, Mumbai 400 005, India.}
\altaffiltext{3}{Instituto Nacional de Astrof\'{\i}sica, \'{O}ptica y Electr\'{o}nica, Luis Enrique Erro \# 1, Tonantzintla, Puebla, M\'{e}xico C.P. 72840.}
\altaffiltext{4}{Institute of Applied Physics of the Russian Academy of Sciences, 46 Ulyanov st., Nizhny Novgorod 603950, Russia.}
\begin{abstract}
In order to investigate star formation (SF) processes in extreme environments, we have carried out a multi-wavelength analysis of the mid-infrared bubble N46, 
which hosts a WN7 Wolf-Rayet (W-R) star. We have used $^{13}$CO line data to trace an expanding shell surrounding the W-R star 
containing about five condensations within the molecular cloud associated with the bubble. 
The W-R star is associated with a powerful stellar wind having a mechanical luminosity of $\sim$4 $\times$ 10$^{37}$  ergs s$^{-1}$. 
A deviation of the H-band starlight mean polarization angles around the bubble has also been traced, indicating the impact of stellar wind on the surroundings. 
The {\it Herschel} temperature map shows a temperature range of $\sim$18--24~K toward the five molecular condensations.  
The photometric analysis reveals that these condensations are associated with the identified clusters of young stellar objects, revealing ongoing SF process. 
The densest among these five condensations (peak N(H$_{2}$) $\sim$9.2 $\times$ 10$^{22}$ cm$^{-2}$ and A$_{V}$ $\sim$98 mag) is associated with a 6.7 GHz methanol maser, an infrared dark cloud, and the CO outflow, 
tracing active massive SF within it. 
At least five compact radio sources (crss) are physically linked with the edges of the bubble and each of 
them is consistent with the radio spectral class of a B0V--B0.5V type star. The ages of the individual infrared counterparts of three crss ($\sim$1--2 Myr) and 
a typical age of WN7 W-R star ($\sim$4 Myr) indicate that the SF activities around the bubble are influenced by the feedback of the W-R star.
\end{abstract}
\section{Introduction}
\label{sec:intro}
Massive stars ($\gtrsim$ 8 M$_{\odot}$) have a significant impact on their surroundings through feedback mechanisms 
such as ionizing radiation, stellar winds, and radiation pressure. 
The W-R stars represent a late phase in the evolution of O stars and can significantly influence the local neighbourhood 
through the kinetic energy contained in their very powerful winds \citep{lamers99}. 
In recent years, infrared emission at {\it Spitzer} 8 $\mu$m band has been used to reveal thousands of bubble/shell-like structures toward the Galactic 
star-forming regions \citep{churchwell06,churchwell07,simpson12}.
Such features often surround the OB stars (including W-R stars), illustrating indirectly their origin due to the energetics of powering source(s).
Such sites have gained considerable interest to explore the interaction between massive stars and their surroundings \citep[e.g.][]{deharveng10}.
Massive stars can stimulate the birth of young stars. 
The feedback of massive stars can also disrupt the parental molecular cloud and thus halt the star formation process.
Together, massive stars can have constructive and destructive effects in their neighbourhood. 
Therefore, the knowledge of physical conditions in such sites can allow to understand 
the star birth process in extreme environments (e.g., regions around OB stars and H\,{\sc ii} regions) 
as well as to infer the predominant feedback mechanism of massive stars. 
Such investigation will enable us to observationally assess the available various theoretical processes ``collect and collapse" 
\citep[see][]{elmegreen77,whitworth94,dale07}, ``radiation-driven implosion" \citep[RDI; see][]{bertoldi89,lefloch94}, and 
the radiation-magnetohydrodynamic simulations of the expansion of the H\,{\sc ii} region around an O star 
in a turbulent magnetized molecular cloud \citep{arthur11}.

The mid-infrared (MIR) bubble, N46 \citep{churchwell06} contains a spectroscopically identified 
W-R 1503-160L star \citep[subtype WN7;][]{shara12} and is a poorly studied region. 
The bubble N46 ($l$ = 27$\degr$.31; $b$ = $-$0$\degr$.11) is referred to as a broken or incomplete ring 
with an average radius (diameter) and thickness of 1$\farcm$38 (2$\farcm$76) 
and 0$\farcm$48 \citep{churchwell06}, respectively. 
The bubble N46 is associated with an H\,{\sc ii} region, G027.3$-$00.1 \citep{lockman89,beaumont10}. 
The line-of-sight velocity of the molecular gas \citep[$\sim$92.6~km\,s$^{-1}$;][]{beaumont10} is consistent with 
the velocity of the ionized gas \citep[$\sim$92.3~km\,s$^{-1}$; recombination linewidth $\sim$32.7~km\,s$^{-1}$;][]{lockman89} 
toward the bubble N46, suggesting the physical association of the molecular and ionized emissions. 
\citet{beaumont10} utilized the CO (J = 3--2) line data around this bubble and 
estimated a near kinematic distance of 4.8$\pm$1.4 kpc, which is adopted throughout the present work.

The detailed and careful investigation of an individual hosting site of the W-R star will allow to explore the star formation process in extreme 
environments. Despite the presence of W-R 1503$-$160L star in the bubble N46, the physical conditions around the bubble are yet to be explored. 
The impact of energetics of W-R 1503$-$160L star on its local environment is not known. 
In this paper, we examine  the distribution of dust temperature, column density, extinction, ionized emission, 
kinematics of molecular gas, distribution of magnetic field lines, 
and young stellar objects (YSOs) around the bubble, using the multi-wavelength data spanning from the centimeter wave 
band to the near-infrared (NIR).  

In Section~\ref{sec:obser}, we describe the multi-wavelength datasets adopted in this work. 
Section~\ref{sec:data} deals with the physical environment around the bubble. 
This section also provides the details of photometric analysis of point-like sources. 
In Section~\ref{sec:disc}, we discuss possible star formation scenarios to explain the presence of young stars around the bubble. 
A summary of conclusions is given in Section~\ref{sec:conc}.
\section{Data and analysis}
\label{sec:obser}
In the present work, we selected a region of $\sim$24$\arcmin$ $\times$ 13$\farcm3$ 
(central coordinates: $l$ = 27$\degr$.3228; $b$ = $-$0$\degr$.1437) around the bubble N46.
\subsection{New Observations}
\subsubsection{Optical Spectrum}
Optical spectrum of W-R 1503-160L star ($l$ = 27$\degr$.2987; $b$ = $-$0$\degr$.1207) was obtained on 2015 April 12 using the Himalaya Faint 
Object Spectrograph and Camera (HFOSC) mounted on the 2 m Himalayan {\it Chandra} Telescope (HCT). 
The spectrum was observed using grism 8 (5800-8350 \AA; R $\sim$2190) of the HFOSC with an exposure time of 30 min. 
A few bias frames and Fe-Ne calibration lamp spectra were also taken for the bias subtraction and calibration of the observed spectrum. 
Following the standard reduction procedures, the observed optical spectrum was extracted using a semi-automated pipeline written in PyRAF \citep{ninan14,baug15}.
The signal-to-noise ratio (S/N) of the optical spectrum is achieved to be 45. 
\subsubsection{Near-infrared Spectra}
The NIR $H$- and $K$-band spectra of W-R star were also observed on 2015 April 13 using the 
TIFR Near Infrared Spectrometer and Imager \citep[TIRSPEC;][]{ninan14} mounted on the HCT. 
Spectra were obtained at two dithered positions with an effective on-source integration time of 300s 
and 600s at $H$- and $K$-bands, respectively. 
TIRSPEC has an average spectral resolution of $\sim$1200. 
Several continuum and Argon lamp spectra were also obtained for the continuum subtraction and wavelength calibration of the observed spectra. 
A separate telluric standard ($\alpha$ Aql) was also observed for telluric correction. Adopting the standard reduction procedures, 
the observed spectra were reduced using a semi-automated pipeline written
 in PyRAF \citep{ninan14,baug15}. Finally, the reduced spectra at $H$ and $K$ bands were flux-calibrated using the 
 Two Micron All Sky Survey \citep[2MASS; resolution $\sim$$2\farcs5$;][]{skrutskie06} photometry of W-R star. 
 The S/N of our observed $H$- and $K$-bands spectra is obtained to be 20 and 45, respectively.
\subsection{Archival data}
To study the star formation process, we used multi-wavelength observations from various Galactic plane surveys 
(e.g. the Multi-Array Galactic Plane Imaging Survey \citep[MAGPIS; $\lambda$ =20 cm;][]{helfand06}, 
the Galactic Ring Survey \citep[GRS; $\lambda$ =2.7 mm;][]{jackson06}, 
the Bolocam Galactic Plane Survey \citep[BGPS; $\lambda$ =1.1 mm;][]{ginsburg13}, 
the APEX Telescope Large Area Survey of the Galaxy \citep[ATLASGAL; $\lambda$ =870 $\mu$m;][]{schuller09}, 
the {\it Herschel} Infrared Galactic Plane Survey \citep[Hi-GAL; $\lambda$ =70, 160, 250, 350, 500 $\mu$m;][]{molinari10}, 
the MIPS Inner Galactic Plane Survey \citep[MIPSGAL; $\lambda$ =24 $\mu$m;][]{carey05}, 
the Wide Field Infrared Survey Explorer \citep[WISE; $\lambda$ =12 $\mu$m;][]{wright10}, 
the Galactic Legacy Infrared Mid-Plane Survey Extraordinaire \citep[GLIMPSE; $\lambda$ =3.6, 4.5, 5.8, 8 $\mu$m;][]{benjamin03}, 
the UKIRT NIR Galactic Plane Survey \citep[GPS; $\lambda$ =1.25, 1.65, 2.2 $\mu$m;][]{lawrence07}, 
the Galactic Plane Infrared Polarization Survey \citep[GPIPS; $\lambda$ =1.6 $\mu$m;][]{clemens12}, 
and 2MASS ($\lambda$ =1.25, 1.65, 2.2 $\mu$m)). 
In the following sections, we give a brief description of these archival data. 

\subsubsection{NIR Data}
We extracted deep NIR photometric {\it JHK} magnitudes of point sources in our selected region from 
the UKIDSS GPS 6$^{th}$ archival data release (UKIDSSDR6plus). 
UKIDSS observations (resolution $\sim$$0\farcs8$) were taken using the UKIRT Wide Field Camera 
\citep[WFCAM;][]{casali07}. 
The final fluxes were calibrated using the 2MASS data. 
One can also obtain more information about the selection procedure of the GPS photometry in \citet{dewangan15}.
Our selected GPS catalog consists of sources fainter than J = 12.0, H = 11.2, and K = 10.0 mag to avoid saturation. 
2MASS data were obtained for bright sources that were saturated in the GPS catalog. 
\subsubsection{H-band Polarimetry}
NIR H-band (1.6 $\mu$m) polarimetric data (resolution $\sim$$1\farcs5$) were downloaded from the GPIPS. 
The data were collected with the {\it Mimir} instrument, on the 1.8 m Perkins telescope, 
in H-band linear imaging polarimetry mode \citep[see][for more details]{clemens12}.
\subsection{{\it Spitzer} Data}
We obtained photometric images and magnitudes of point sources in our selected region from 
the {\it Spitzer}-GLIMPSE survey at 3.6--8.0 $\mu$m (resolution $\sim$2$\arcsec$). 
The data were retrieved from the GLIMPSE-I Spring '07 highly reliable catalog. The image at MIPSGAL 24 $\mu$m was collected. 
We also used MIPSGAL 24 $\mu$m photometry \citep[from][]{gutermuth15} in this work. 
\subsection{{\it Herschel}, ATLASGAL, and BGPS Data}
{\it Herschel} far-infrared and sub-millimeter(mm) continuum maps at 70 $\mu$m, 160 $\mu$m, 250 $\mu$m, 350 $\mu$m, and 500 $\mu$m were obtained 
from the {\it Herschel} data archive. The beam sizes of these images are 5$\farcs$8, 12$\arcsec$, 18$\arcsec$, 25$\arcsec$, and 37$\arcsec$ \citep{poglitsch10,griffin10}, respectively. The processed level2$_{-}$5 products were downloaded using the {\it Herschel} Interactive Processing 
Environment \citep[HIPE,][]{ott10}.

The sub-mm continuum map at 870 $\mu$m (beam size $\sim$19$\farcs$2) was retrieved from the ATLASGAL. 
Bolocam 1.1 mm image was also obtained from the BGPS. 
The beam size of the 1.1 mm map is $\sim$33$\arcsec$. 
\subsection{$^{13}$CO (J=1$-$0) Line Data}
The $^{13}$CO (J=1$-$0) line data were obtained from the GRS. 
The GRS data have a velocity resolution of 0.21~km\,s$^{-1}$, an angular resolution 
of 45$\arcsec$ with 22$\arcsec$ sampling, a main beam efficiency ($\eta_{\rm mb}$) of $\sim$0.48, 
a velocity coverage of $-$5 to 135~km~s$^{-1}$, and a typical rms sensitivity (1$\sigma$)
of $\approx0.13$~K \citep{jackson06}.  
\subsection{Radio Centimeter Continuum Map}
Radio continuum map at 20 cm was extracted from the VLA MAGPIS. 
The map has a 6\farcs2 $\times$ 5\farcs4 beam and a pixel scale of 2$\arcsec$/pixel.
\section{Results}
\label{sec:data}
\subsection{Optical and NIR spectra of W-R star}
\label{sec:spect}
\begin{figure*}
\epsscale{1.0}
\plotone{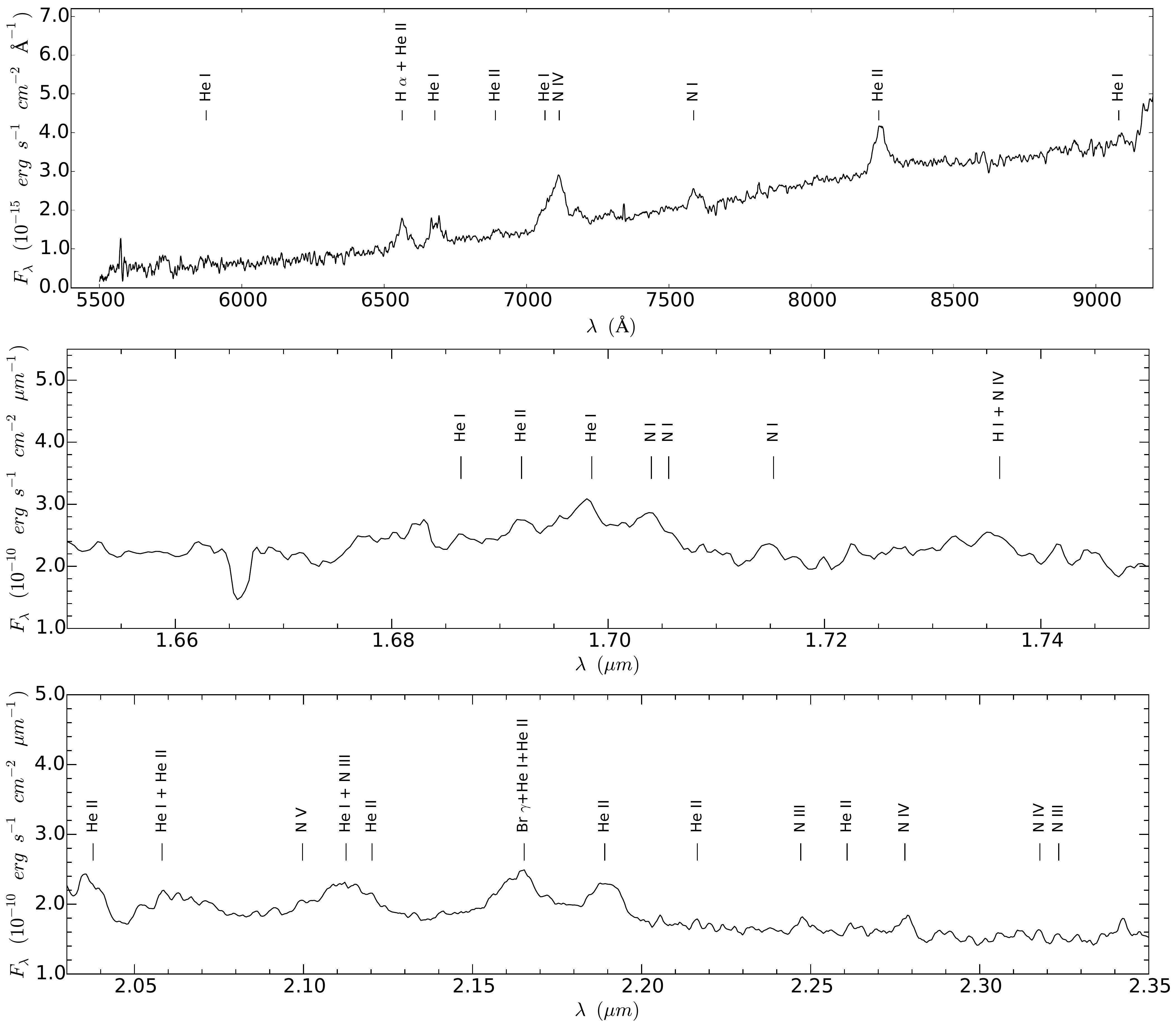}
\caption{\scriptsize Optical (top panel), H-band (middle panel), and K-band (bottom panel) spectra of W-R 1503-160L 
star ($l$ = 27$\degr$.2987; $b$ = $-$0$\degr$.1207; the location of the W-R star is shown in Figure~\ref{fig1}) (see the text). 
The broad emission lines of ionized helium and nitrogen are seen (also see Table~\ref{tab2aa}).}
\label{fig1aa}
\end{figure*}

Spectroscopic observations offer a unique opportunity to characterize the individual astronomical objects. 
In Figure~\ref{fig1aa}, we present optical (5500--9200 \AA) and NIR $H$- and $K$-band spectra of W-R 1503-160L star. 
The optical and NIR spectra mainly show the broad emission lines of ionized helium and nitrogen, confirming it as W-R star. 
In general, due to strong stellar wind of W-R star, the outer hydrogen envelopes are thrown away and 
thus, strong lines of He and other heavy elements (N, O, C) are observed in spectra of W-R stars. 
In the present work, we have utilized both optical and NIR spectra to determine the spectral type of this source. 
\citet{smith96} have demonstrated a classification scheme of W-R stars using the helium lines (HeII at 4686 and 5411 \AA, and HeI 
at 5875 \AA) in the optical spectrum. 
 \citet{figer97} have also shown that several characteristic lines in the NIR $K$-band spectrum 
 can be used to determine the spectral subtype of W-R star. 
 Our optical spectrum does not cover the lines in the blue part, as used by \citet{smith96}, hence, 
 we primarily carried out the identification of the spectral type using our 
 $K$-band spectrum. Absence of carbon lines and presence of highly ionized nitrogen lines in all the spectra
 preliminarily confirm the source as a nitrogen-rich W-R star (e.g. WN subtype) (see Table~\ref{tab2aa}). 
 From our observed $K$-band spectrum, the equivalent widths of 2.112 $\mu m$ (HeI + NIII), 2.166 $\mu m$ (HI + HeI + HeII), and 2.189 (HeII) lines were calculated by fitting Gaussian profiles (see Table~\ref{tab2aa}). Good agreement of the ratio of equivalent widths (2.189/2.112 $\sim$0.85 and 2.189/2.166 $\sim$0.72) corresponding to WN7 star, as reported in Figure~1 of \citet{figer97}, confirms the source as WN7 type star. 
 Furthermore, we have compared the ratio of equivalent widths of several optical lines with those values reported in \citet{conti90}.
 The ratios of equivalent-widths of HeI 6678 \AA~and HeII 8237 \AA~(ratio $\sim$ 1.27), and HeI 6678 \AA~and 
 blended HeI 7065 \AA~+ NIV 7715 \AA~(ratio $\sim$ 0.37) agree well with the values 
 reported for several WN7 stars (e.g., WR 55, WR 120, WR 148, and WR 158) in \citet{conti90}. 
 Hence, from our observed optical and NIR spectra we confirm the source as WN7 W-R star. 
 Note that the source was previously characterized as WN7 W-R type star using the {\it K}-band 
 spectrum (with a spectral resolution of $\sim$1200) by \citet{shara12}. 
 Our K-band spectrum (with a spectral resolution of $\sim$1000) is similar to the spectrum reported in \citet{shara12}. 
 Similar characteristics of broad emission lines of He and N are seen in the present study as well as previously published work. 
 We have given equivalent widths of our observed lines in Table~\ref{tab2aa}, which are consistent with the NIR lines listed in Table~5 in \citet{shara12}. 
 They utilized the 2MASS colors of the source and estimated an extinction to be $\sim$0.84 mag in K$_{s}$-band. 
 They also computed a distance to W-R 1503-160L star of $\sim$3.1 kpc, assuming it is a strong-lined WN7 star.

In this work, we have also utilized the widths of lines seen in our observed spectra and have estimated the terminal velocity of the emitting gas of W-R star.
In the first step, we determined the instrumental broadening of the lines using the isolated observed sky emission lines, 
which is found to be $\sim$430 km s$^{-1}$. 
Furthermore, the equivalent widths of  HeII + H$\alpha$ line at 6563 \AA~and Br$\gamma$ + HeI + HeII line at 2.166 $\mu$m 
were used to determine the velocity of the 
expanding gas.  Finally, after correcting the velocity for the instrumental broadening, we obtain the terminal 
velocity of the outflowing gas to be $\sim$1600$\pm$400 km s$^{-1}$. 

\citet{crowther06} extensively studied the properties of W-R stars in a super cluster, Westerlund 1, and 
suggested a criterion to infer the strong-lined and weak-lined WN7 W-R stars. 
They mentioned that the strong-lined WN7 W-R stars have full width at half 
maximum (FWHM) of HeII 2.1885 $\mu m$ line of $\ge$130 \AA. 
They also listed the absolute magnitudes of strong-lined and weak-lined WN7 W-R stars, 
which are different for these two subclasses. 
The absolute magnitudes (in K$_{s}$ band) of strong-lined and weak-lined WN7 W-R stars were reported to be $-$4.77 mag and $-$5.92 mag, 
respectively \citep[see Table~A1 in][]{crowther06}, which will lead two different estimates of distance. 
Taking into account these absolute magnitudes along with 2MASS photometry (m$_{K_{s}}$ = 8.51 mag) and 
extinction (A$_{K_{s}}$ $\sim$0.84 mag; as mentioned above) in K$_{s}$-band, we compute the distances 
to strong-lined and weak-lined WN7 W-R 1503-160L stars of 3.1$\pm$0.7 kpc and 5.2$\pm$0.7 kpc, respectively. 
Using our $K$-band spectra, the FWHM of HeII 2.1885 $\mu m$ line of W-R 1503-160L is about 110 \AA, 
which implies the source to be a weak-lined WN7 star. Considering this result and our calculations, we adpot the distance 
to weak-lined WN7 W-R star of 5.2$\pm$0.7 kpc, which is in good agreement with the kinematic distance of the bubble N46. 
Hence, taken together, we suggest that the bubble N46 hosts the W-R 1503-160L star.
\subsection{Physical environment around the bubble N46}
\label{sec:env}
The knowledge of the ionized, dust, and molecular emissions in a given star-forming complex allows to infer physical conditions, 
and to provide details about H\,{\sc ii} regions, possible outflows, and photon dominated regions. 
Additionally, the distribution of molecular gas in a given star-forming region also helps us to trace its exact physical extension/boundary. 
\subsubsection{Global View}
\begin{figure*}
\epsscale{0.6}
\plotone{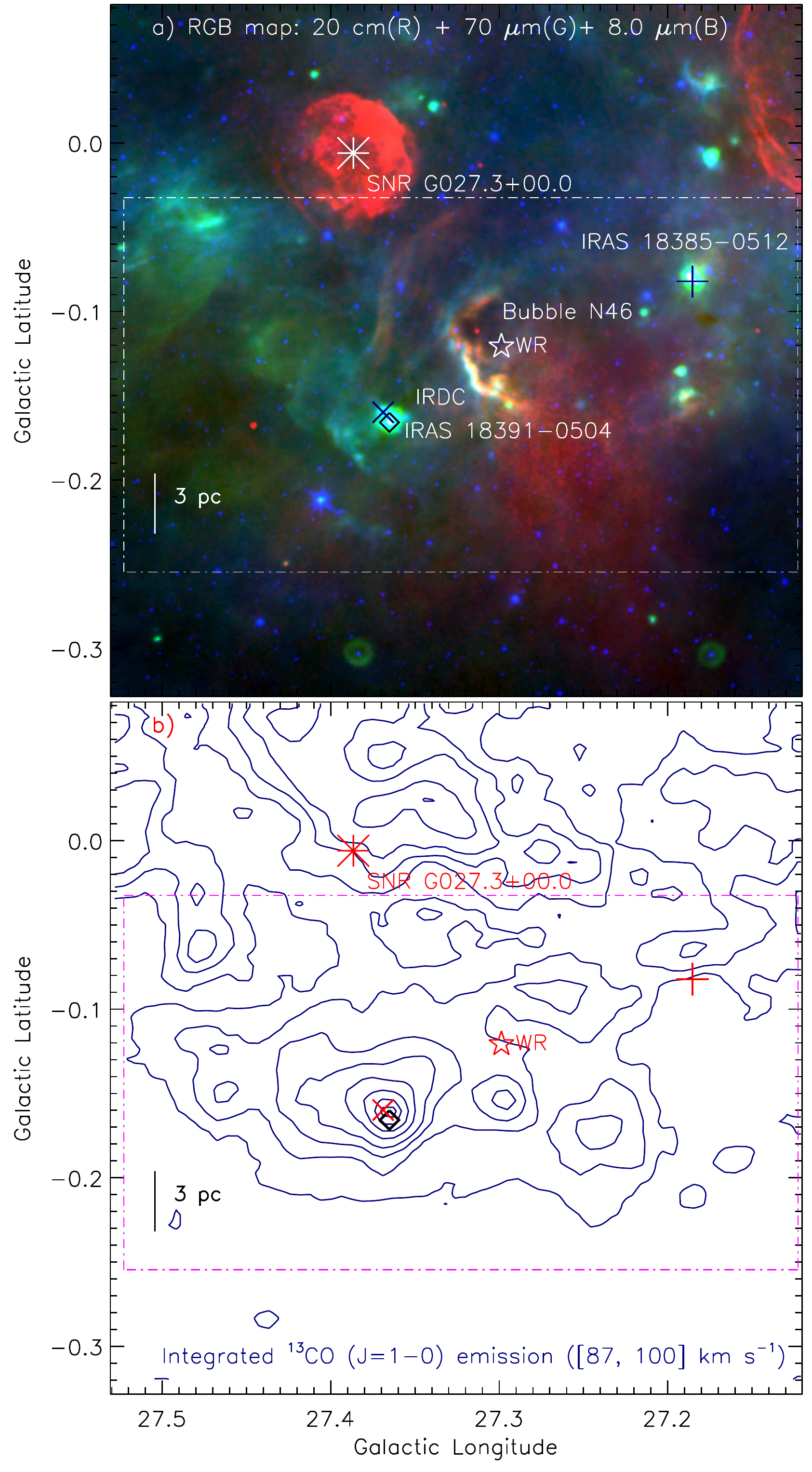}
\caption{\scriptsize A global view of the region around the bubble N46 (size of the selected field $\sim 24\farcm6  \times 24\farcm6$; central
coordinates: $l$ = 27$\degr$.3260; $b$ = $-$0$\degr$.1236). 
a) The image is the result of the combination of three bands (in linear scale): 20 cm in red (MAGPIS), 70 $\mu$m in green ({\it Herschel}), 
and 8.0 $\mu$m in blue ({\it Spitzer}). 
The color composite map depicts the ionized emission (in 20 cm map), warm dust emission (in 70 $\mu$m image), 
and PAH emission including the continuum (in 8 $\mu$m image).
 b) Molecular $^{13}$CO gas in the direction of the bubble N46. 
The CO integrated velocity range is from 87 km s$^{-1}$ to 100 km s$^{-1}$. 
The CO contours are 45.27 K km s$^{-1}$ $\times$ (0.1, 0.2, 0.3, 0.4, 0.55, 0.7, 0.85, 0.95).
In both panels, the positions of IRAS 18385$-$0512 (+), IRAS 18391$-$0504 ($\times$), 
SNR G027.3+00.0 ($\ast$), a 6.7-GHz methanol maser emission ($\Diamond$), and a W-R 1503$-$160L star ($\star$) 
are marked. The scale bar corresponding to 3 pc at a kinematical distance of 4.8 kpc is shown in both panels. 
In both panels, the dotted-dashed box shows the field of Figure~\ref{fig2}.}
\label{fig1}
\end{figure*}

In Figure~\ref{fig1}, we show the spatial distribution of ionized emission, warm dust emission, 
and molecular gas around the bubble N46. 
Figure~\ref{fig1}a is a three-color composite image made using MAGPIS 20 cm in red, {\it Herschel} 70 $\mu$m in green, 
and GLIMPSE 8.0 $\mu$m in blue. Radio continuum map at MAGPIS 20 cm reveals several ionized 
regions. The image at 70 $\mu$m represents the warm dust emission, whereas the 8 $\mu$m band is dominated 
by the 7.7 and 8.6 $\mu$m polycyclic aromatic hydrocarbon (PAH) emission (including the continuum). 
In Figure~\ref{fig1}b, we show the molecular $^{13}$CO (J = 1--0) gas emission in the direction 
of the bubble N46. The $^{13}$CO profile depicts the bubble region in a velocity range of 87--100 km s$^{-1}$. 
SNR G027.3+00.0 (also known as Kes 73) is seen as the most prominent ionized region in the radio map (see Figure~\ref{fig1}a). 
The radio map also traces some compact point-like radio sources. 
The composite map also displays the edges of the bubble as a prominent bright feature coincident with the 20 cm emission (see Figure~\ref{fig1}a). 
In Figure~\ref{fig1}a, we have marked the positions of some known astronomical 
objects (see Table~\ref{tab2} for coordinates), which are seen in the global view of the bubble N46. 
As mentioned before, the bubble harbors WN7 W-R 1503-160L star that is situated at a projected linear separation of 
$\sim$8$\farcm$67 from the SNR G027.3+00.0. 
\citet{ferrand12} tabulated the distance and age of the SNR G027.3+00.0 to be 7.5--9.1 kpc 
and 750--2100 year, respectively. 
Using photometric and spectroscopic analyses (as mentioned earlier), we find the distance of W-R 1503-160L star to be 
5.2$\pm$0.7 kpc, which is consistent with the distance of the bubble (i.e. 4.8$\pm$1.4 kpc). 
In general, W-R stars have very high mass-loss rates (2--5 $\times$ 10$^{-5}$ M$_{\odot}$/yr) with terminal velocities of 1000--2500 km s$^{-1}$ \citep{prinja90} 
and have life time roughly a few million years \citep{lamers99}. 
Considering the distance of the bubble N46, these two sources, SNR G027.3+00.0 and bubble N46, do not appear to be physically linked. 
Therefore, the impact of the SNR G027.3+00.0 on the surroundings of the bubble N46 is unlikely. 
\subsubsection{Local environment}
\begin{figure*}
\epsscale{0.94}
\plotone{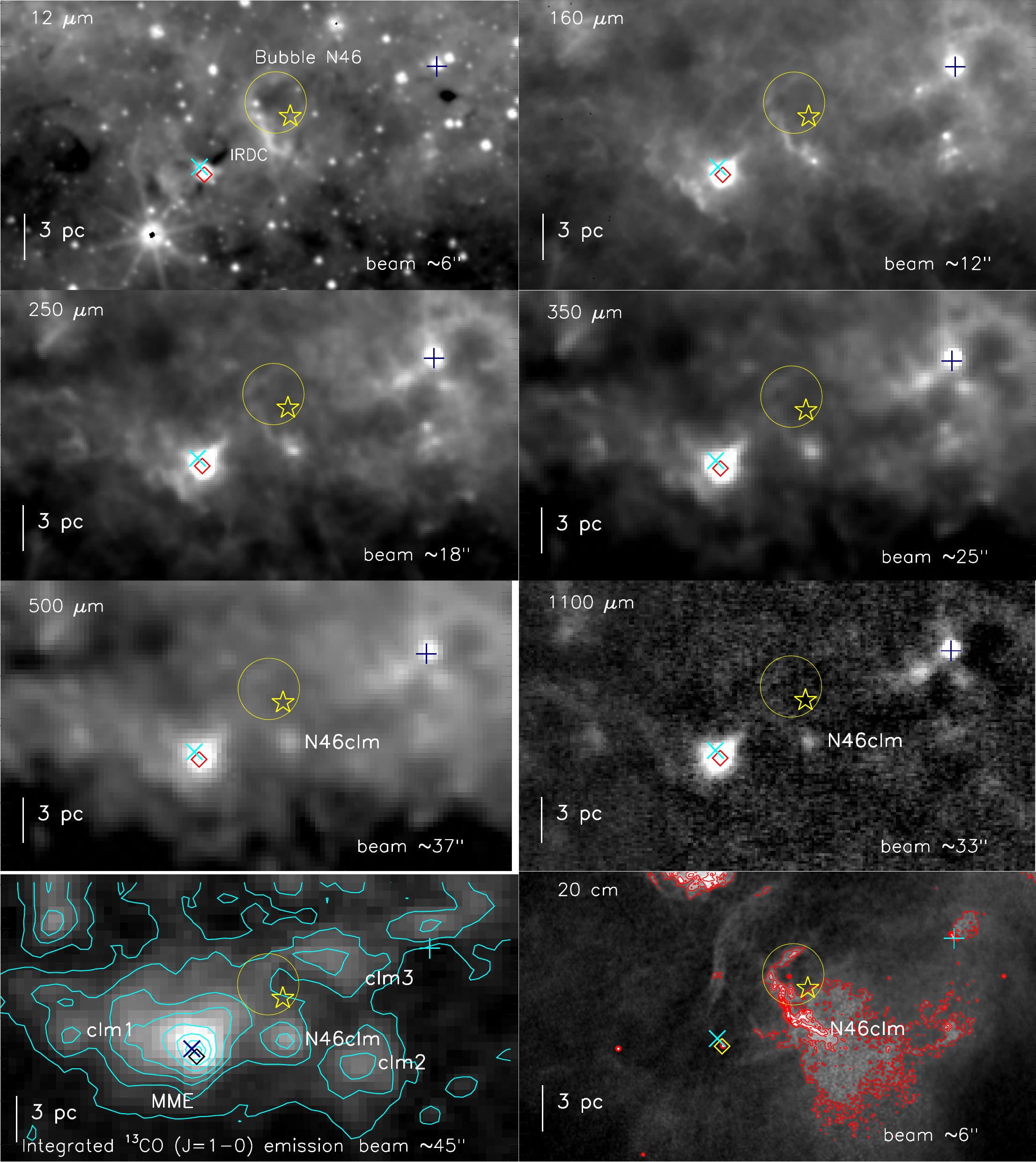}
\caption{\scriptsize The distribution of warm dust, cold dust, molecular gas, and ionized emissions in the region around the bubble N46, 
using the WISE, {\it Herschel}, BOLOCAM, GRS, and MAGPIS surveys. 
The selected area (size of the region $\sim 24$\arcmin$ \times 13\farcm3$; central coordinates: $l$ = 27$\degr$.3228; $b$ = $-$0$\degr$.1437) 
is shown by a dotted-dashed box in Figure~\ref{fig1}a. 
The panels display images at 12 $\mu$m, 160 $\mu$m, 250 $\mu$m, 350 $\mu$m, 500 $\mu$m, 
1100 $\mu$m, integrated $^{13}$CO emission, and 20 cm, from left to right in increasing order. 
The CO emission is also shown by cyan contours with 45.18 K km s$^{-1}$ $\times$ (0.1, 0.2, 0.3, 0.4, 0.55, 0.7, 0.85, 0.95). 
In the integrated CO map, some noticeable peak 
positions (i.e. clm1, clm2, clm3, MME, and N46clm) are labeled. 
The MAGPIS 20 cm emission is also overlaid by red solid contours with levels of 0.0023, 0.0035, 0.0046, 
0.0064, 0.0081, 0.0099, and 0.011 Jy/beam. In all the panels, the other marked symbols are similar to those shown in Figure~\ref{fig1}. 
The position of the W-R 1503$-$160L star ($\star$) is marked in all the panels. 
In all the panels, a big circle indicates the location of the bubble N46 (with a diameter of $\sim$2$\farcm$76 or 
3.85 pc at a distance of 4.8 kpc) as reported by \citet{churchwell06}.} 
\label{fig3} 
\end{figure*}

\begin{figure*}
\epsscale{1.0}
\plotone{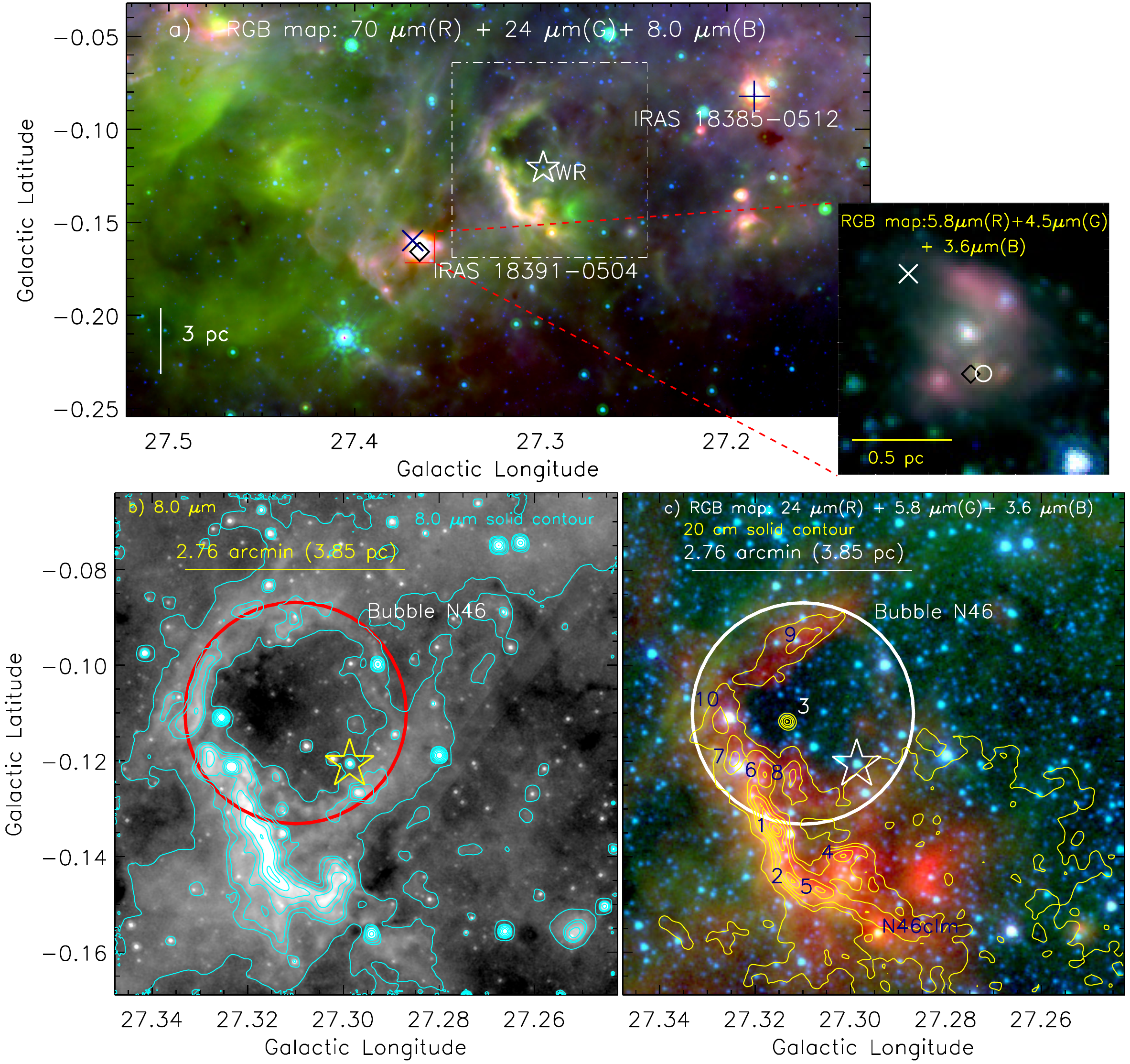}
\caption{\scriptsize a) Composite color image (70 $\mu$m (red), 24 $\mu$m (green), and 8.0 $\mu$m (blue); in log scale) of the 
region around the bubble N46. 
The selected area is shown by a dotted-dashed box in Figure~\ref{fig1}a. 
The inset on the bottom right represents the IRAS 18391$-$0504 region in zoomed-in view, using 
a three-color-composite image made of 
three {\it Spitzer} images (5.8 $\mu$m (red), 4.5 $\mu$m (green), and 3.6 $\mu$m (blue)) in log scale (see the solid red box in figure). 
A small white circle represents the peak position of 5 GHz detection in the inset panel.
The highlighted dotted-dashed box in figure shows the field of Figures~\ref{fig2}b and~\ref{fig2}c. 
b) A zoomed-in view of the bubble N46 using {\it Spitzer} 8 $\mu$m gray-scale image (in log scale). 
The 8 $\mu$m contours are also overlaid with levels of 1.46, 1.65, 1.7, 1.8, 1.9, 2.0, and 2.1 MJy/sr (in log scale). 
c) A close-up view of the bubble N46 using three {\it Spitzer} images (24 $\mu$m (red), 5.8 $\mu$m (green), and 3.6 $\mu$m (blue); in log scale). 
The MAGPIS 20 cm contours in yellow are similar to those shown in Figure~\ref{fig3}. 
The compact radio sources (crss) are highlighted by numbers in blue color (also see Table~\ref{fig3}). 
In the last two panels, a thick big circle shows the location of the bubble N46. 
In all the panels, the other marked symbols are similar to those shown in Figure~\ref{fig1}.}
\label{fig2}
\end{figure*}

In Figure~\ref{fig1}b, based on the distribution of molecular emission, 
we have selected the region around the bubble for our present study, 
where majority of molecular gas is distributed (hereafter, N46 molecular cloud; see dotted-dashed box in Figure~\ref{fig1}b).  
In Figure~\ref{fig3}, we show the multi-wavelength images of the selected region around the bubble N46 at 12 $\mu$m, 160 $\mu$m, 250 $\mu$m, 350 $\mu$m, 500 $\mu$m, 1100 $\mu$m, integrated $^{13}$CO emission, and 20 cm. 
The radio continuum and integrated CO maps are also shown for comparison with the infrared and sub-mm images. 
The details of integrated CO map as well as kinematics of molecular gas are given in Section~\ref{sec:coem}.
We have highlighted the positions of IRAS 18385$-$0512, IRAS 18391$-$0504, 
6.7-GHz methanol maser emission (MME), and W-R star. WISE 12 $\mu$m band traces the warm dust emission and 
contains PAH emission feature at 11.3 $\mu$m. The 160--1100 $\mu$m images indicate the presence of the cold dust emission 
(see Section~\ref{subsec:temp} for quantitative estimates). 
We identify about five molecular condensations (e.g. clm1, clm2, clm3, MME, and N46clm) based on the visual inspection 
of the integrated molecular map, which are labeled in the map. The MME condensation is found to be associated with 
an infrared dark cloud (IRDC) in the 8 $\mu$m image. 
The IRDC appears as a bright in emission at wavelengths longer than 24 $\mu$m. 
In Figure~\ref{fig2}a, we show a color composite image made using 8 $\mu$m (blue), 24 $\mu$m (green), and 70 $\mu$m (red) images.  
In Figure~\ref{fig2}a, the inset on the bottom right shows the zoomed-in view toward the MME and 
IRAS 18391$-$0504 (a color composite image: 5.8 $\mu$m (red), 4.5 $\mu$m (green), and 3.6 $\mu$m (blue) images). 
The observed MME is located $\sim$25$\farcs$3 away 
from the IRAS 18391$-$0504 (see inset in Figure~\ref{fig2}a). 
In Figures~\ref{fig2}b and~\ref{fig2}c, we present the zoomed-in view of the bubble N46, which depict the edges of the bubble.
Figure~\ref{fig2}c shows the absence of radio continuum peak toward W-R star located inside the bubble. 
Note that the 20 cm continuum emission is lacking inside the bubble except a detection of a point-like radio source. 
Additionally, the 24 $\mu$m emission, traces the warm dust emission, is not enclosed by the 8 $\mu$m emission. 
It indicates that the bubble structure is unlikely originated by the expanding H\,{\sc ii} region.

\citet{shirley13} reported the molecular HCO$^{+}$ (3--2) gas velocities to be $\sim$91.1 and $\sim$89.2 km s$^{-1}$ toward 
the N46clm and clm1 condensations, respectively. The peak velocity of 6.7-GHz MME is $\sim$99.7 km s$^{-1}$ \citep{szymczak12}. 
Ammonia (NH$_{3}$) line observations were also reported toward MME condensation by \citet{wienen12}. 
They found the radial velocities of the NH$_{3}$ gas between 91.21 and 91.70 km s$^{-1}$ and also estimated the near kinematic distance as 5.06 kpc. 
However, the line-of-sight velocity of molecular gas toward the IRAS 18385$-$0512 is $\sim$26 km s$^{-1}$ \citep{sridharan02}. 
With the knowledge of these velocities, it appears that the bubble N46, W-R star, and five condensations 
(including IRAS 18391$-$0504 and 6.7-GHz MME) belong to the same region (see Figure~\ref{fig2}b), however, the IRAS 18385$-$0512 is not physically associated with the N46 molecular cloud. 
Considering this fact, we do not discuss the results related to IRAS 18385$-$0512 in the present work. 

All together, with the help of molecular line data and distribution of cold, warm, and ionized emission, 
we find the physical association of the W-R star and molecular condensations located inside the N46 molecular cloud. 
Furthermore, the ionized emission appears to be well correlated with the warm dust emission at the edges of the bubble. 
\subsubsection{Compact radio sources}
In Figure~\ref{fig2}c, the radio peaks (or compact radio source (crs)) are seen toward the edges of the bubble and are designated by numbers as 1-10 (also see Table~\ref{tab3}). 
In this section, the number of Lyman continuum photon (N$_{uv}$) is computed for each crs 
using the integrated flux density following the equation of \citet{matsakis76} \citep[see][for more details]{dewangan15a}. 
We used the ``CLUMPFIND" IDL program \citep{williams94} to obtain the integrated flux density for each crs. 
The calculations were performed for a distance of 4.8 kpc and for the electron temperature of 10000~K.
The values of integrated flux density and N$_{uv}$ are listed in Table~\ref{tab3}. The table also contains the spectral class of each of the crss 
(see Table 1 in \citet{smith02} for theoretical values). Each crs appears to be associated with a single ionizing star of 
spectral type B0V/B0.5V (see Table~\ref{tab3}). 
\subsubsection{Infrared counterpart of IRc and MME}
In Figure~\ref{fig2}c, we have examined the infrared counterpart (IRc) of each crs (within a radius of 4$\arcsec$ of the radio peak) 
via a visual inspection of GLIMPSE images.
We find the IRcs of crs7 (designated as G027.3235$-$00.1212), crs9 (designated as G027.3115$-$00.0948), and crs10 (designated as G027.3255$-$00.1109). 
The analysis of the NIR color-magnitude diagram also suggests these sources as massive stars earlier than spectral type B0 
(see Section~\ref{subsec:sed} for spectral energy distribution (SED) modeling). 

Additionally, we identify the embedded sources G027.2942$-$00.1559, G027.4598$-$00.1509, and G027.3650$-$00.1656 toward the 
peaks of N46clm, clm1, and MME condensations, respectively. 
The source G027.2942$-$00.1559 is detected at wavelengths longer than 2.2~$\mu$m, whereas the source G027.4598$-$00.1509 is seen only 
at wavelengths longer than 3.6 $\mu$m. 
We find an IRc of 6.7-GHz MME within a radius of 4$\arcsec$, which is designated as G027.3650$-$00.1656 in the GLIMPSE catalog and 
has GLIMPSE 4.5 and 5.8 $\mu$m bands photometry (m$_{4.5}$ = 10.07$\pm$0.3 mag 
and m$_{5.8}$ = 8.35$\pm$0.2 mag). 
\citet{purcell13} reported the Coordinated Radio and Infrared Survey for High-Mass Star Formation \citep[CORNISH;][]{hoare12} 
5~GHz (beam $\sim$1$\farcs$5) radio source (i.e. G027.3644$-$00.1657 (angular scale $\sim$1$\farcs$7)) close to 
this IRc (separation of $\sim$2$\farcs$1; see inset panel in Figure~\ref{fig2}a). They also estimated the integrated flux density equal to 60.14 mJy, 
which refers to log(N$_{uv}$) of $\sim$47.0 s$^{-1}$ and corresponds to a single powering star of radio spectral class B0.5V. 
A compact point-like radio continuum source at MAGPIS 20 cm is also detected toward the 6.7-GHz MME.  
In general, the detection of 6.7-GHz MME strongly indicates the presence of early phases of massive star formation (MSF) \citep[e.g.][]{walsh98,urquhart13}.

In summary, the multi-wavelength images reveal the presence of MSF (B0V type stars and 6.7 GHz MME) activities around the bubble.
\subsection{Kinematics of molecular gas}
\label{sec:coem} 
\begin{figure*}
\epsscale{1}
\plotone{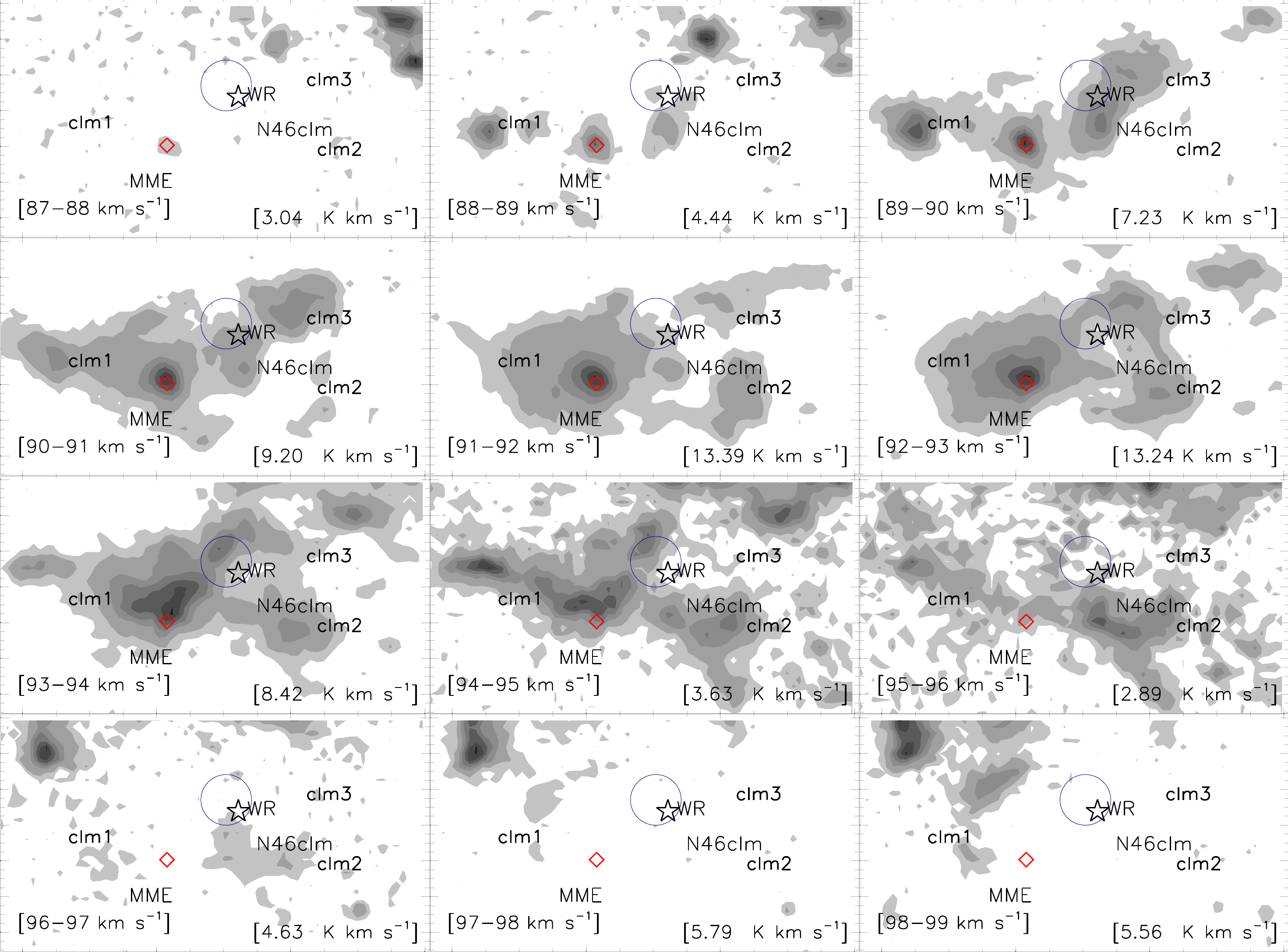}
\caption{\scriptsize $^{13}$CO(J =1$-$0) velocity channel contour maps. 
The velocity value (in km s$^{-1}$) is shown in each panel. 
The contour levels are 10, 20, 40, 55, 70, 85, and 98\% of the peak value (in K km s$^{-1}$), which is also indicated in each panel. 
Other marked symbols and labels are similar to those shown in Figure~\ref{fig3}.}
\label{fig666}
\end{figure*}

\begin{figure*}
\epsscale{1.0}
\plotone{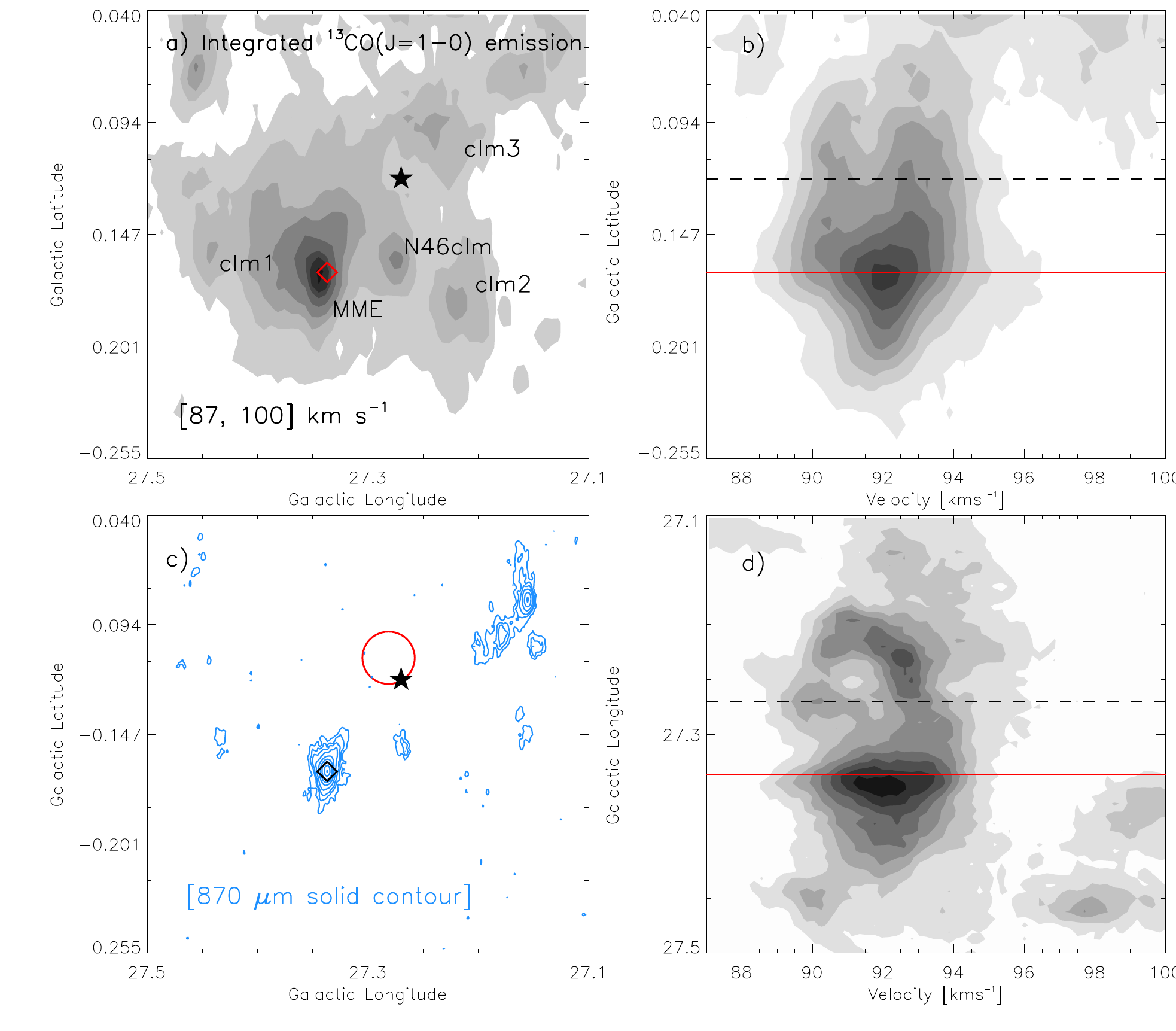}
\caption{\scriptsize  a) Integrated intensity map of $^{13}$CO (J = 1-0). 
The map is similar to the one shown in Figure~\ref{fig3}.
b) Latitude-velocity distribution of $^{13}$CO. The CO emission is integrated over the longitude from 27$\degr$.1 to 27$\degr$.5.
c) The ATLASGAL 870 $\mu$m contours (in dodger blue) are shown with levels of 
10.24 Jy/beam $\times$ (0.015, 0.03, 0.045, 0.08, 0.15, 0.25, 0.4, 0.7, and 0.98). 
The bubble location is similar to the one shown in Figure~\ref{fig2}b. 
d) Longitude-velocity distribution of $^{13}$CO. The CO emission is integrated over the latitude from $-$0.$\degr$04 to $-$0.$\degr$255. 
In both the left panels, the positions of the 6.7-GHz MME ($\Diamond$) and the W-R 1503$-$160L star ($\star$) 
are marked. Solid red line and dashed black line show the positions of the 6.7-GHz MME and the W-R 1503$-$160L 
star in the position-velocity maps, respectively. In panel d, the plot depicts the inverted C-like morphology 
as well as a noticeable velocity gradient toward the 6.7-GHz MME (also see the text).}
\label{fig7}
\end{figure*}

This section deals with a kinematic analysis of the molecular gas in the N46 molecular cloud. 
The kinematics of the molecular gas may help us to explore the expansion of the bubble. 
Figure~\ref{fig666} shows the integrated GRS $^{13}$CO (J=1$-$0) velocity channel 
maps (at intervals of 1 km s$^{-1}$), revealing different molecular condensations along the line of sight. 
The channel maps show a clear separation of the high- and low-velocity emission toward the MME condensation. 
In the velocity range from 88 to 95 km s$^{-1}$, the highlighted molecular condensations (i.e. clm1, clm2, clm3, MME, and N46clm) are well detected. Using GRS $^{13}$CO (J=1$-$0) and radio continuum data, \citet{anderson09} examined the 
properties of molecular cloud associated with the Galactic H\,{\sc ii} regions including the bubble N46, 
which is referred as C27.31$-$0.14 (V$_{lsr}$ $\sim$92.3 km s$^{-1}$) in their catalog. The integrated GRS $^{13}$CO intensity 
map\footnote[1]{http://www.bu.edu/iar/files/script-files/research/hii\_regions/region\_pages/C27.31-0.14.html} 
was provided by these authors, however, the position-velocity analysis of this cloud is still lacking. 
In Figure~\ref{fig7}, we present the integrated $^{13}$CO intensity map and the position-velocity maps. 
The galactic position-velocity diagrams of the $^{13}$CO emission reveals an almost inverted C-like 
structure as well as the presence of a noticeable velocity gradient toward the 6.7 GHz MME (see Figures~\ref{fig7}b and~\ref{fig7}d).  
In Figure~\ref{fig7}c, we show the bubble location traced at 8 $\mu$m along with the distribution of cold dust emission. 
As previously mentioned, the 6.7 GHz MME represents MSF and is associated with the densest clump in the N46 molecular cloud. 
Additionally, the MME condensation contains a cluster of young populations (see Section~\ref{subsec:surfden}). 
The position-velocity diagrams convincingly suggest the presence of an outflow activity toward the MME condensation. 
In the channel maps, the orientations of the high- and low-velocity lobes appear significantly different 
(see the velocity range between 88 to 95 km s$^{-1}$ in Figure~\ref{fig666}). 
These features suggest the presence of multiple CO outflows in the MME condensation. 
We cannot further explore this aspect in this work due to the coarse beam of the GRS CO data (beam size 45$\arcsec$). 
The existence of inverted C-like structure in the position-velocity plot often indicates an expanding shell \citep{arce11}.
\citet{arce11} examined the observed molecular gas distribution in the Perseus molecular cloud 
along with a modeling of expanding bubbles in a turbulent medium. 
They indicated that the semi-ring-like or C-like structure in the position-velocity plot 
represents an expanding shell. The position of the W-R star is approximately situated at the center of the inverted C-like structure. 
Our position-velocity analysis indicates the presence of an expanding shell with an 
expansion velocity of the gas to be $\sim$3 km s$^{-1}$.

Taken together, the CO kinematics suggest the presence of at least one outflow as well as the expanding shell. 
\subsection{{\it Herschel} temperature and column density maps}
\label{subsec:temp}
In this section, we study the distribution of column density, temperature, extinction, and clump mass in the N46 molecular cloud.
We followed the procedures given in \citet{mallick15} to obtain the {\it Herschel} temperature and column density maps (see below).  
We obtained these maps from a  pixel-by-pixel SED fit with a modified blackbody curve to the cold 
dust emission in the {\it Herschel} 160--500 $\mu$m wavelengths. 
The images at 250--500 $\mu$m are in the surface brightness unit of MJy sr$^{-1}$, while the image at 160 $\mu$m is 
calibrated in the units of Jy pixel$^{-1}$. The plate scales of the 160, 250, 350, and 500 $\mu$m images are 6.4, 6, 10, and 14 arcsec/pixel, respectively. 

Here, we provide a brief step-by-step description of the procedures. 
All images were first converted to Jy pixel$^{-1}$ unit and were convolved to the angular resolution of 500 $\mu$m image ($\sim$37$\arcsec$), 
using the convolution kernels available in the HIPE software. The images were regridded  to the pixel size of 
500 $\mu$m image ($\sim$14$\arcsec$) using the HIPE software.
Next, the sky background flux level was measured to be  1.476, 3.705, 1.759, and 0.643 Jy pixel$^{-1}$ for the 160, 250, 350, and 
500 $\mu$m images (size of the selected region $\sim$6$\farcm$6 $\times$ 7$\farcm$3; 
central coordinates: $\alpha_{J2000}$ = 18$^{h}$44$^{m}$45$^{s}$, $\delta_{J2000}$ = -05$\degr$40$\arcmin$00$\arcsec$), respectively.
The featureless dark region far from the N46 molecular cloud, to avoid diffuse emission associated with target, was carefully chosen for the background estimation.   

\begin{figure*}
\epsscale{0.97}
\plotone{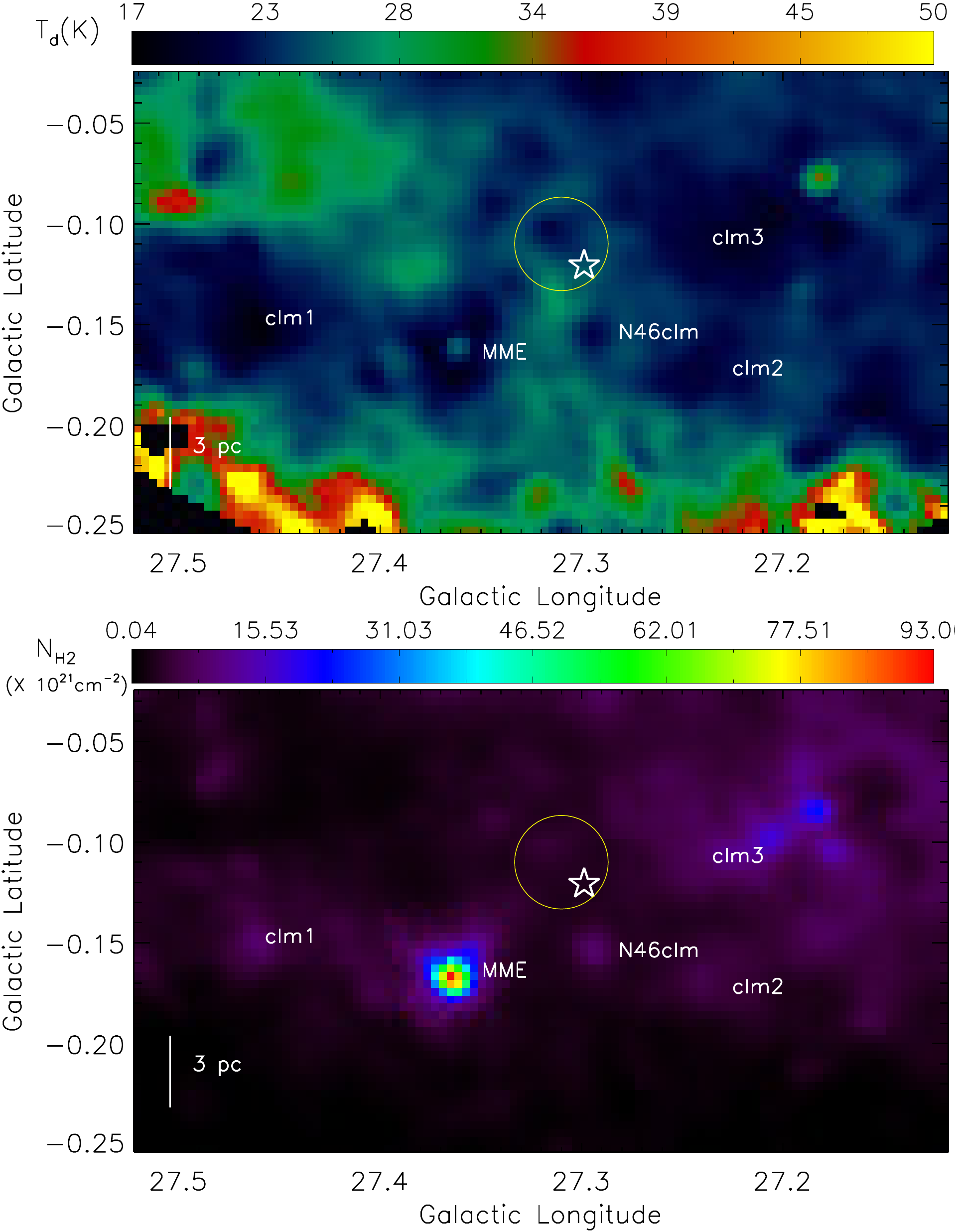}
\caption{\scriptsize {\it Herschel} temperature map (top panel) and column density ($N(\mathrm H_2)$) map (bottom panel) 
of our selected region around the bubble N46 (see text for details). 
The extinction value can be inferred from the column density 
using the relation: $A_V=1.07 \times 10^{-21}~N(\mathrm H_2)$. 
In the temperature map, the black area in a latitude range between $-$0.25$\degr$ and $-$0.2$\degr$ corresponds to NaN values. 
In both panels, the other marked symbols and labels are similar to those shown in Figure~\ref{fig3}.}
\label{fig4}
\end{figure*}

Finally, we fitted a modified blackbody to the observed fluxes on a pixel-by-pixel basis to obtain the maps \citep[see equations 8 and 9 given in][]{mallick15}. 
We performed the fitting using the four data points for each pixel, retaining the 
dust temperature (T$_{d}$) and the column density ($N(\mathrm H_2)$) 
as free parameters. 
In the estimations, we used a mean molecular weight per hydrogen molecule ($\mu_{H2}$=) 2.8 
\citep{kauffmann08} and an absorption coefficient ($\kappa_\nu$ =) 0.1~$(\nu/1000~{\rm GHz})^{\beta}$ cm$^{2}$ g$^{-1}$, 
including a gas-to-dust ratio ($R_t$ =) of 100, with a dust spectral index of $\beta$\,=\,2 \citep[see][]{hildebrand83}. 
In Figure~\ref{fig4}, we show the final temperature and column density maps (resolution $\sim$37$\arcsec$) of our 
selected region around the bubble N46. In Figure~\ref{fig4}a, we find warmer gas (T$_{d}$ $\sim$23-27 K) toward the edges of the bubble. 
{\it Herschel} temperature map also exhibits a temperature range of about 18--24~K toward the five molecular condensations as marked in Figure~\ref{fig3}. 
We find the peak column densities of about 9.2~$\times$~10$^{21}$ (A$_{V}$ $\sim$10 mag), 
6.31~$\times$~10$^{21}$ (A$_{V}$ $\sim$7 mag), 1.18~$\times$~10$^{22}$ (A$_{V}$ $\sim$12 mag), 
9.2~$\times$~10$^{22}$ (A$_{V}$ $\sim$98 mag), and 9.17~$\times$~10$^{21}$ (A$_{V}$ $\sim$10 mag) cm$^{-2}$ toward the molecular condensations clm1, clm2, clm3, MME, and N46clm, respectively (see Figure~\ref{fig4}b). 
Here, we used the relation between optical extinction and hydrogen column density from \citet{bohlin78} (i.e. $A_V=1.07 \times 10^{-21}~N(\mathrm H_2)$). 
Note that the clump associated with 6.7 GHz MME is relatively warmer (T$_{d}$ $\sim$24 K) and is the densest one compared to other condensations.
Using NH$_{3}$ line observations, \citet{wienen12} reported physical parameters of dense gas (i.e. the gas kinetic temperature 
and rotational temperature) toward the MME clump. 
The kinetic temperature of the MME clump was estimated to be $\sim$23.81 K, which is consistent with our estimated temperature value. 

We estimate the masses of four prominent clumps (MME, N46clm, clm1, and clm2) seen in the column density map (see Figure~\ref{fig4}b). 
The mass of a single clump can be obtained using the formula:
\begin{equation}
M_{clump} = \mu_{H_2} m_H Area_{pix} \Sigma N(H_2)
\end{equation}
where $\mu_{H_2}$ is assumed to be 2.8, $Area_{pix}$ is the area subtended by one pixel, and 
$\Sigma N(\mathrm H_2)$ is the total column density. 
We employed the ``CLUMPFIND" program to estimate the total column densities of the clumps. 
Using equation~1, we compute the masses of the 
clumps MME, N46clm, clm1, and clm2 to be $\sim$7273, $\sim$954, $\sim$412, and $\sim$741 $M_\odot$, respectively.
\subsection{Young stellar objects in the N46 molecular cloud}
\subsubsection{Selection of infrared excess sources}
\label{subsec:phot1}
\begin{figure*}
\epsscale{0.98}
\plotone{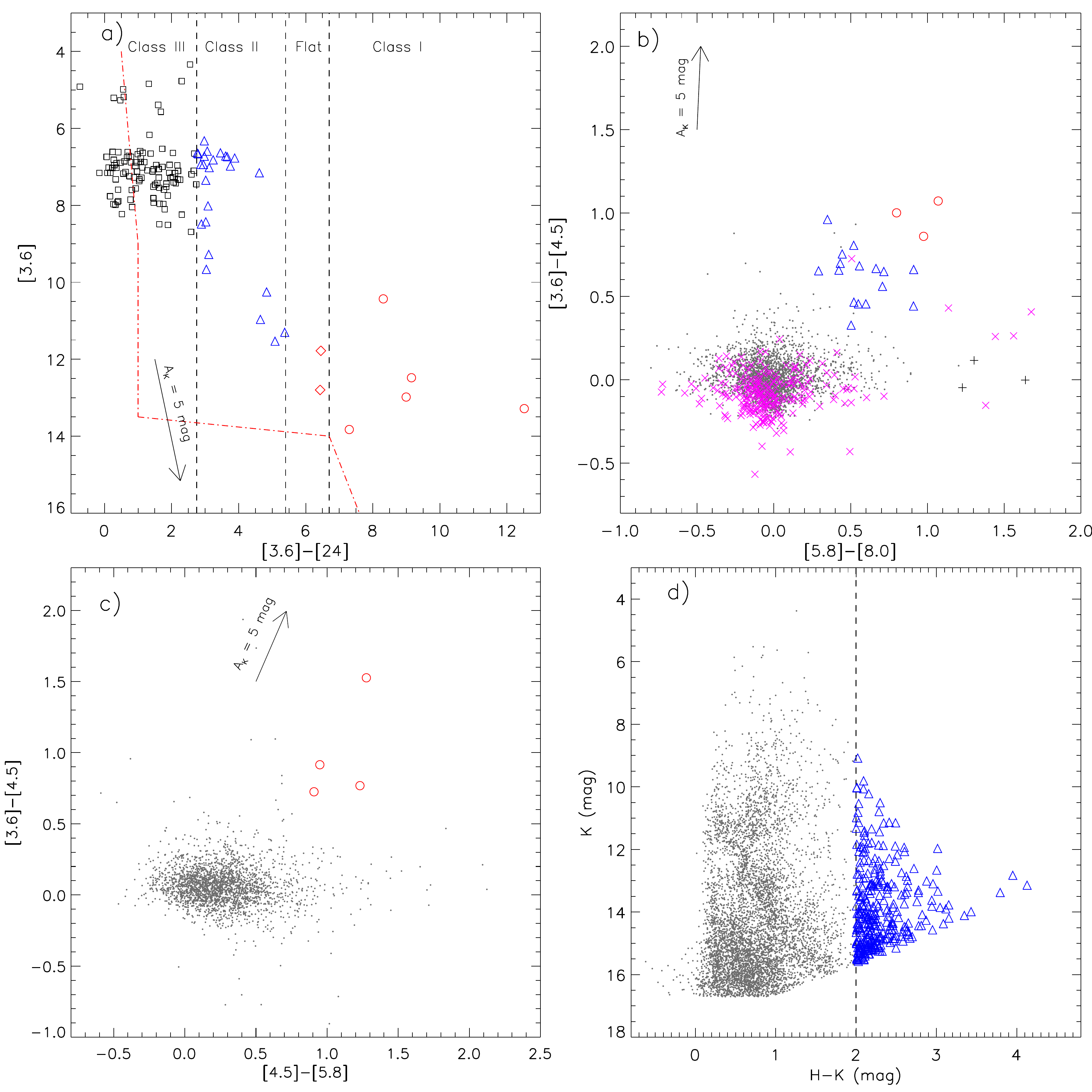}
\caption{\scriptsize Identification of young stellar population within our selected region around the bubble N46 (see Figure~\ref{fig2}). 
a) Color-magnitude diagram ([3.6] $-$ [24] vs [3.6]) of sources observed in the IRAC and 
MIPSGAL bands (see the text for more details). 
The diagram allows to distinguish YSOs belonging to different evolutionary stages (see dashed lines). 
The dotted-dashed lines represent the separation chosen for YSO's against contaminated 
candidates (galaxies and disk-less stars) \citep[see][for more details]{rebull11}. 
The ``$\Diamond$'' and ``$\Box$'' symbols represent the Flat-spectrum and Class~III sources, respectively; 
b) Color-color diagram ([3.6]$-$[4.5] vs. [5.8]$-$[8.0]) using the IRAC four band detections. 
The ``+'' and ``$\times$'' symbols represent the PAH-emitting galaxies and PAH-emission-contaminated apertures, respectively (see the text); 
c) Color-color diagram ([3.6]$-$[4.5] vs. [4.5]$-$[5.8]) of the sources detected in three IRAC bands, except 8.0 $\mu$m image; 
d) Color-magnitude diagram (H$-$K/K) of the sources detected in the NIR bands. 
In all the panels, we show Class~I (red circles) and Class~II (open blue triangles) YSOs. 
In the last three panels, the dots in gray color show the stars with only photospheric emissions. 
In the NIR H$-$K/K plot, we have plotted only 6001 out of 45909 stars with photospheric emissions. 
In the first three panels, the arrow represents the extinction vector corresponding to the average extinction laws from \citet{flaherty07}.}
\label{fig5}
\end{figure*}

The study of YSOs is considered a primary tracer of the star-formation activity in a given star-forming region. 
In the following, we use different photometric surveys to identify and classify YSOs.\\\\
1. We selected sources having detections in MIPSGAL 24 $\mu$m and GLIMPSE 3.6 $\mu$m bands.
The color-magnitude plot ([3.6] $-$ [24]/[3.6]) has been utilized to trace the youngest population in a given star-forming 
region \citep{guieu10,rebull11,dewangan15}. In our selected region, we find 135 sources that are common in the 3.6 and 24 $\mu$m bands. 
In Figure~\ref{fig5}a, we present the [3.6] $-$ [24]/[3.6] plot of 135 sources.
We select 32 YSOs (5 Class I; 2 Flat-spectrum; 25 Class~II) and 103 Class~III sources. 
The boundary of different stages of YSOs is also marked in the Figure~\ref{fig5}a \citep[e.g.][]{guieu10}, 
which depicts the criteria to infer the youngest population.  
The figure also shows the boundary of possible contaminants (i.e. galaxies and disk-less stars) \citep[see Figure~10 in][]{rebull11}.
We do not identify any contaminants (i.e. galaxies) in our selected YSOs. \\

2. We obtained sources having detections in all four GLIMPSE bands.
Following the \citet{gutermuth09} schemes, we identified YSOs and various possible contaminants (e.g. broad-line active galactic nuclei (AGNs), 
PAH-emitting galaxies, shocked emission blobs/knots, and PAH-emission-contaminated apertures) in our selected region around the bubble.
The possible contaminants are also removed from the selected YSOs. 
We further utilized the slopes of the {\it Spitzer}-GLIMPSE SED ($\alpha_{IRAC}$) measured from 3.6 to 8.0 $\mu$m \citep[e.g.,][]{lada06} and 
classified the selected YSOs into different evolutionary stages (i.e. Class~I ($\alpha_{IRAC} > -0.3$), Class~II ($-0.3 > \alpha_{IRAC} > -1.6$), 
and Class~III ($-1.6> \alpha_{IRAC} > -2.56$)). One can also find more details of YSO classifications in \citet[][and references therein]{dewangan11}. 
In Figure~\ref{fig5}b, we show the GLIMPSE color-color diagram ([3.6]$-$[4.5] vs [5.8]$-$[8.0]) for all the identified sources. 
In the end, we find 19 YSOs (3 Class I; 16 Class~II), 2545 photospheres, 3 PAH-emitting galaxies, 
and 235 contaminants.\\ 

3. Infrared excess traced in the first three GLIMPSE bands (except 8.0 $\mu$m band) can also offer to identify 
the additional protostars. The color-color space ([4.5]$-$[5.8] vs [3.6]$-$[4.5]) is utilized to trace the infrared excess sources. 
We use color conditions, [4.5]$-$[5.8] $\ge$ 0.7 and [3.6]$-$[4.5] $\ge$ 0.7, to select protostars, as given in \citet{hartmann05} and \citet{getman07}. We identify 4 protostars in our selected region around the bubble (see Figure~\ref{fig5}c). \\ 

4. The NIR H-K color excess also allows to identify additional YSOs. In order to choose a color criterion, we examined 
the color-magnitude (H$-$K/K) plot of the nearby control field 
(size $\sim$6$\arcmin$  $\times$ 6$\arcmin$; central coordinates: $\alpha_{J2000}$ = 18$^{h}$41$^{m}$53$^{s}$.6, 
$\delta_{J2000}$ = -05$\degr$14$\arcmin$55$\arcsec$.5) and obtained a color H$-$K value (i.e. $\sim$2.0) that 
isolates large H$-$K excess sources from the rest of the population. 
Adopting this color H$-$K cut-off condition, we obtained 315 embedded YSOs in our 
selected region around the bubble (see Figure~\ref{fig5}d).\\

Finally, using all the four schemes mentioned above, we obtain a total of 370 YSOs in our 
selected region around the bubble (as shown in Figure~\ref{fig2}a). The positions of all selected YSOs are shown in Figure~\ref{fig6}a.
\subsubsection{Study of distribution of YSOs}
\label{subsec:surfden}
\begin{figure*}
\epsscale{1.0}
\plotone{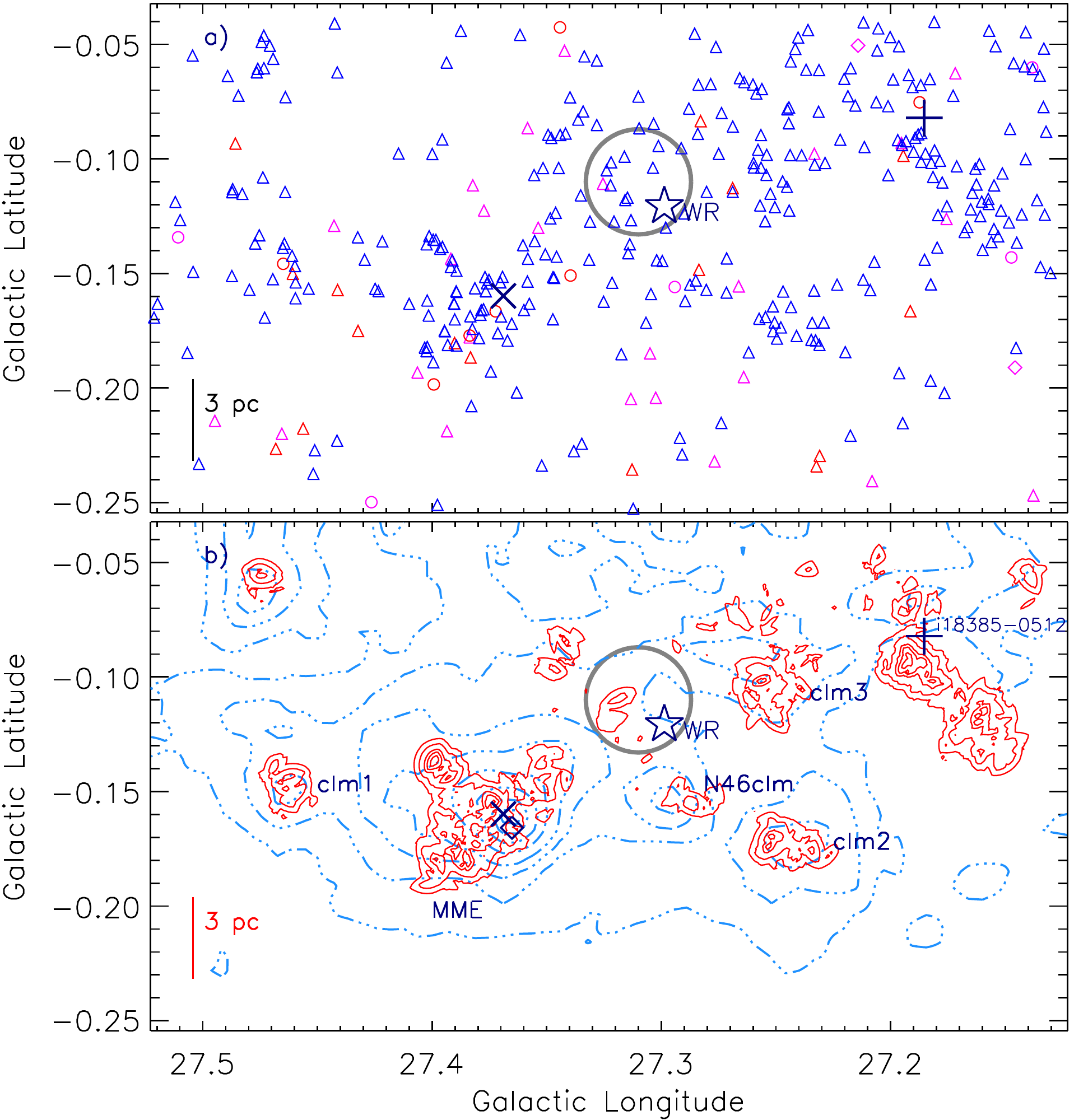}
\caption{\scriptsize a) The positions of YSOs (Class~I (circles), Flat-spectrum (diamond), and Class~II (triangles)) identified within our selected region. 
The YSOs selected using the IRAC-MIPSGAL scheme (see Figure~\ref{fig5}a) are shown by magenta color, 
while the YSOs identified using the other schemes (see the last three panels in Figure~\ref{fig5}) are marked in red (Class~I) and blue (Class~II) colors.
b) The surface density contour map (in red) of all the identified YSOs within our selected region. 
The surface density contours are shown at 2$\sigma$ (1.5 YSOs/pc$^{2}$, where 1$\sigma$=0.74 YSOs/pc$^{2}$), 
3$\sigma$ (2 YSOs/pc$^{2}$), 5$\sigma$ (4 YSOs/pc$^{2}$), 7$\sigma$ (5 YSOs/pc$^{2}$), and 11$\sigma$ (8 YSOs/pc$^{2}$), increasing from the outer to the inner regions. The CO emission is also drawn by cyan contours similar to those shown in Figure~\ref{fig3}. 
In both panels, the other marked symbols and labels are similar to those shown in Figure~\ref{fig3}. 
The figure depicts the association of molecular gas and clusters of YSOs.}
\label{fig6}
\end{figure*}

In this section, we utilized the standard surface density analysis to study the spatial distribution of our selected YSOs \citep[see][for more details]{casertano85,gutermuth09,bressert10,dewangan11,dewangan15}. 
The surface density analysis is a useful tool to reveal the individual groups or clusters of YSOs.
Following the procedure and equation given in \citet{dewangan15}, the surface density map of YSOs was 
generated using the nearest-neighbour (NN) technique. 
The map was obtained using a 5$\arcsec$ grid and 6 NN at a distance of 4.8 kpc. 
In Figure~\ref{fig6}b, we present the resultant surface density contours of YSOs and also highlight the bubble location in the figure. 
The surface density contour levels are drawn at 2$\sigma$ (1.5 YSOs/pc$^{2}$, where 1$\sigma$=0.74 YSOs/pc$^{2}$), 
3$\sigma$ (2 YSOs/pc$^{2}$), 5$\sigma$ (4 YSOs/pc$^{2}$), 7$\sigma$ (5 YSOs/pc$^{2}$), 
and 11$\sigma$ (8 YSOs/pc$^{2}$), increasing from the outer to the inner regions. 
We find noticeable YSOs clusters toward all the five molecular condensations (clm1, clm2, clm3, MME, and N46clm; see Figure~\ref{fig6}b). 
The study of distribution of YSOs clearly illustrates the star formation activity in the N46 molecular cloud (see Figure~\ref{fig6}). 
Note that the surface density contours of YSOs seen toward the IRAS 18385$-$0512 do not appear to 
be associated with the bubble N46 (see Section~\ref{sec:env}). 
\subsubsection{The SED modeling}
\label{subsec:sed}
\begin{figure*}
\epsscale{1.0}
\plotone{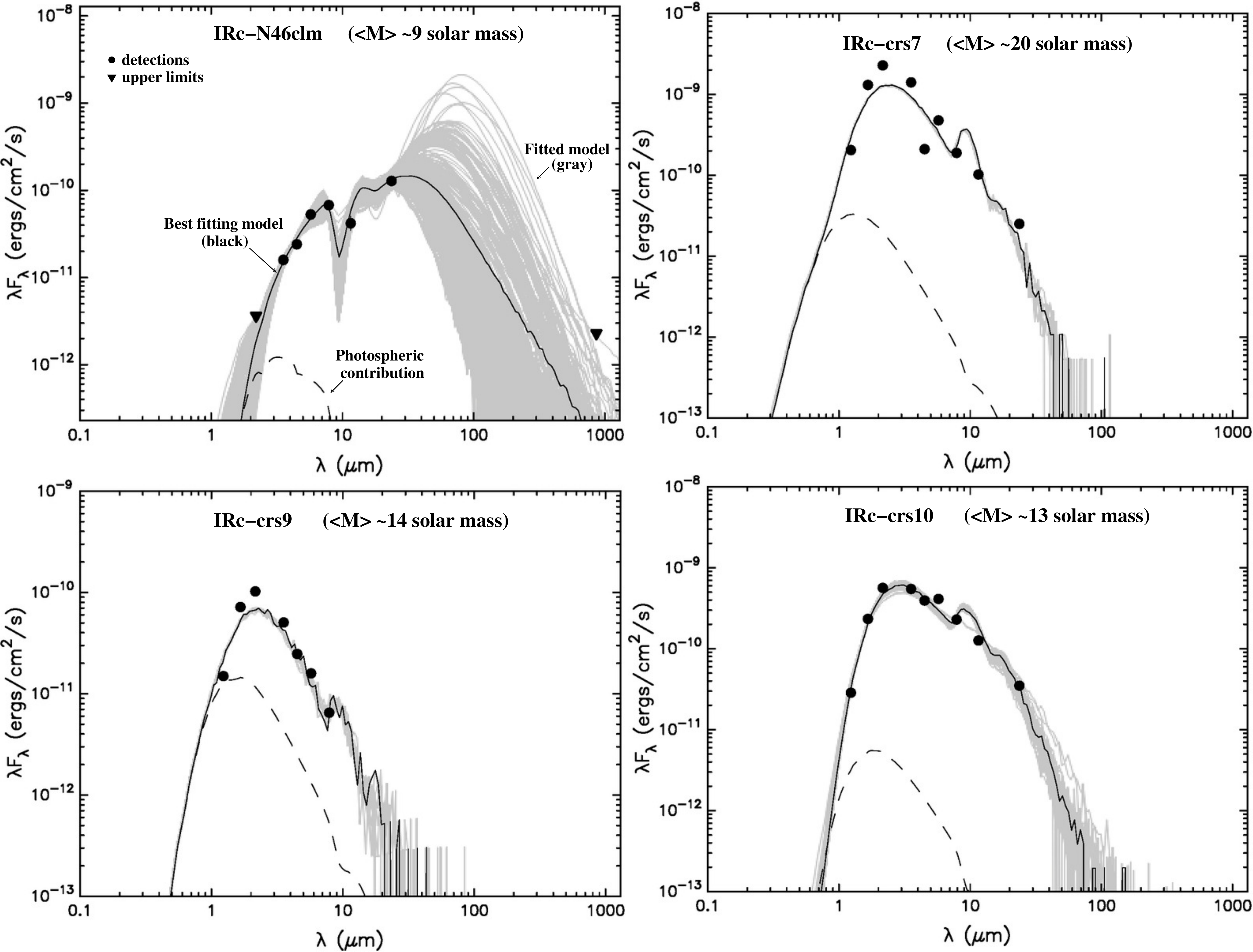}
\caption{\scriptsize The SED modeling of IRcs of N46clm, crs7, crs9, and crs10 (also see Figure~\ref{fig2}c) using the \citet{robitaille06} 
fitting technique. 
Only those models are shown which satisfy the condition ($\chi^{2}$ - $\chi^{2}_{best}$) per data point $<$ 3. 
The black solid line represents the best fit; the gray solid curves show all other models giving a good fit to the data. 
The dashed curve highlights photospheric contribution.} 
\label{fig8gg} 
\end{figure*}

In this section, we performed the SED modeling of IRcs of crs7, crs9, crs10, and N46clm (see Section~\ref{sec:env}) using an 
on-line SED fitting tool \citep{robitaille06,robitaille07}. 
The SED fitting tool needs a minimum of three data points with good quality and fits the data allowing the 
distance to the source and a range of visual extinction values as free parameters. 
The accretion scenario is incorporated into the SED models, which assume a central source associated with 
rotationally flattened infalling envelope, bipolar cavities, and a flared accretion disk. 
The model grid encompasses a total of 200,000 SED models \citep{robitaille06}, 
covering a range of stellar masses from 0.1 to 50 M$_{\odot}$. The SED fitting tool fits only those models which satisfy the 
criterion $\chi^{2}$ - $\chi^{2}_{best}$ $<$ 3, where $\chi^{2}$ is taken per data point. 
The resultant fitted SED models of these IRcs are presented in Figure~\ref{fig8gg}. 
Reliable photometric magnitudes of IRcs have been used for modeling, which can be seen as data points in the fitted SED plots. 
We provided the visual extinction in the range 0--100 mag, as an input parameter for the SED modeling. 
The weighted mean values of the stellar mass (extinction) of IRcs of crs7, crs9, crs10, and N46clm 
are 20$\pm$0.1 M$_{\odot}$ (6.8$\pm$0.1 mag), 14$\pm$0.6 M$_{\odot}$ (9.4$\pm$0.1 mag), 13$\pm$1 
M$_{\odot}$ (13.5$\pm$1.5 mag), and 9$\pm$3 M$_{\odot}$ (54$\pm$19 mag), respectively. 
The weighted mean values of the stellar age of IRcs of crs7, crs9, crs10, and N46clm 
are estimated to be 1.2 Myr, 1.3 Myr, 1.8 Myr, and 0.67 Myr, respectively.
Note that the IRcs of crs10 and N46clm are identified as YSOs. 
Our SED results are consistent with inferred radio spectral types of the crss and 
further suggest the presence of massive stars on the edges of the bubble. 
\subsection{GPIPS H-band Polarization}
\label{subsec:pol}
In the following, we study the large scale morphology of the plane-of-the-sky projection of 
the magnetic field toward the bubble. 
The polarization of background starlight is a useful tool to trace  the  projected 
plane-of-the-sky magnetic field morphology. 
The polarization vectors of background stars give the field direction in the plane of the sky parallel to the direction of
polarization \citep{davis51}.

\begin{figure*}
\epsscale{0.93}
\plotone{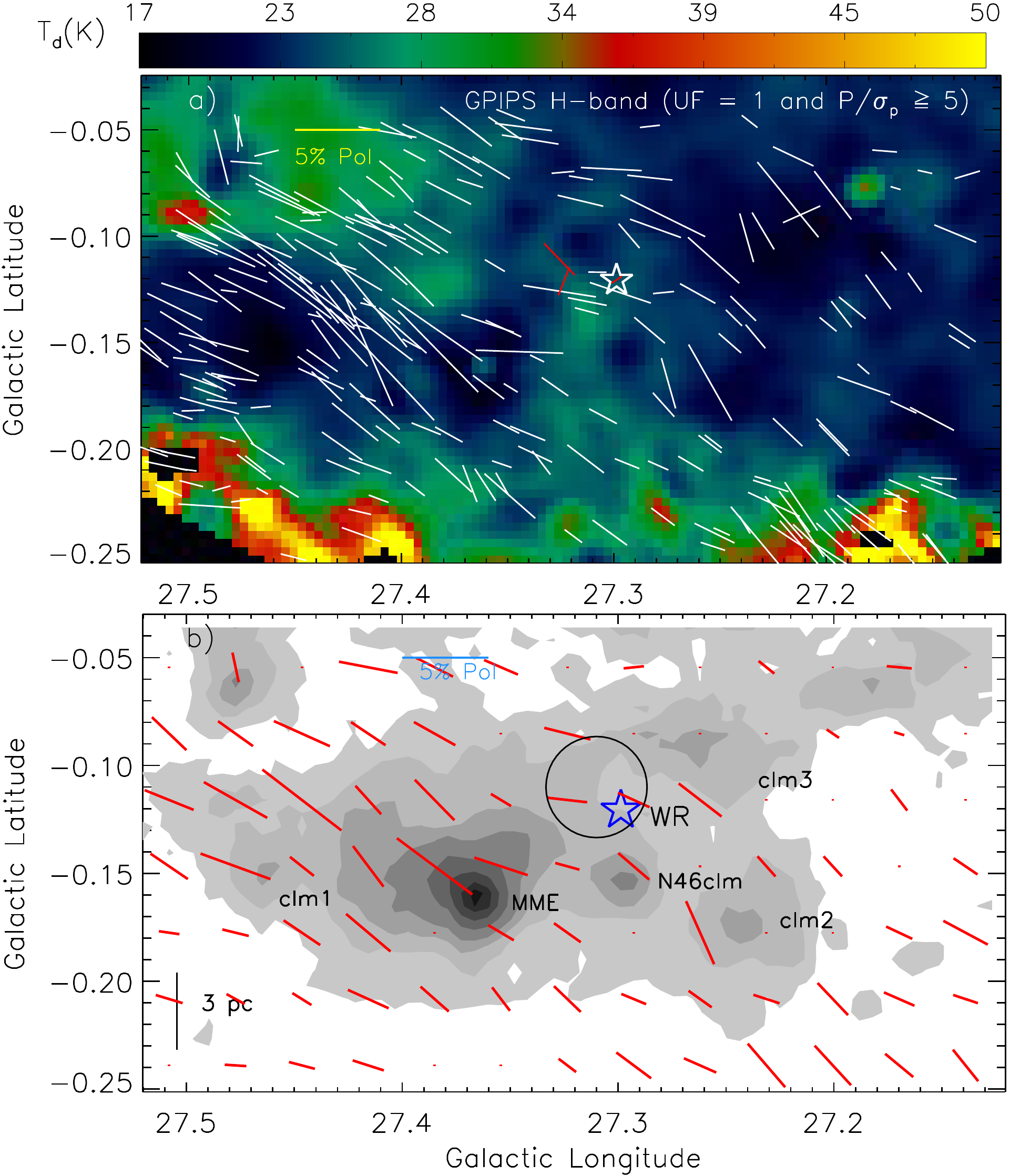}
\caption{\scriptsize a) {\it Herschel} temperature map is overlaid with the GPIPS H-band polarization vectors (in white) of 249 
stars with UF = 1 and $P/\sigma_p \ge$ 5. The H-band polarization vectors of crs7, crs10, and W-R stars are shown in red color. 
The background map is similar to the one shown in Figure~\ref{fig4}a. 
b) The integrated $^{13}$CO emission map is overplotted with the mean polarization vectors. 
The background map is similar to the one shown in Figure~\ref{fig3}. 
The mean polarization data are computed by dividing the polarization spatial region into 13 $\times$ 7 equal parts 
and, a mean polarization value of H-band sources is obtained within each specific part. 
The marked symbols and labels are similar to those shown in Figure~\ref{fig3}. 
In both panels, the degree of polarization is inferred from the length of each vector. 
The orientations of the vectors show the galactic position angles of polarization in both panels. 
In both panels, the ``star" symbol indicates the location of the W-R star. A reference vector of 5\% is drawn in both panels.}
\label{fig8}
\end{figure*}

In Figure~\ref{fig8}a, we display the H-band polarization vectors overlaid on the {\it Herschel} temperature map. 
The polarization data toward our selected region around the bubble were retrieved from the 
GPIPS \citep[see][for more details]{clemens12} and were covered in five fields i.e., 
GP0768, GP0769, GP0782,  GP0783, GP0796, and GP0797. In order to obtain the reliable polarimetric information, 
we selected sources with Usage Flag (UF) = 1 and $P/\sigma_p \ge$ 5. Following these conditions, we obtain a total of 249 stars.
We also find the H-band polarization of crs7, crs10, and W-R stars. These sources have suffered extinction, as listed in previous sections. 
Therefore, it appears that the H-band polarization could be related to the sources and may not be associated with the diffuse interstellar medium. 
Hence, we have shown the polarization vectors of these sources with an addition of 90$\degr$ in their respective position angles (see red vectors in Figure~\ref{fig8}a). 
To examine the distribution of H-band polarization, we show mean polarization vectors in Figure~\ref{fig8}b. 
In order to infer the mean polarization, our selected spatial area is divided into 13 $\times$ 7 equal divisions 
and a mean polarization value is computed using the Q and U Stokes parameters of H-band sources 
located inside each specific division. 
In Figure~\ref{fig8}, the degree of polarization is traced by the length of a vector, whereas the angle of a vector indicates 
the polarization galactic position angle. 
Note that the H-band polarization data cannot allow us to infer the morphology of the plane-of-the-sky projection of the magnetic field toward the 
dense clumps (clm1, clm2, clm3, MME, and N46clm). 

The mean distribution of H band vectors appears uniform, however we notice a little change in the mean polarization pattern 
between the MME condensation and  W-R star. 

\begin{figure*}
\epsscale{0.65}
\plotone{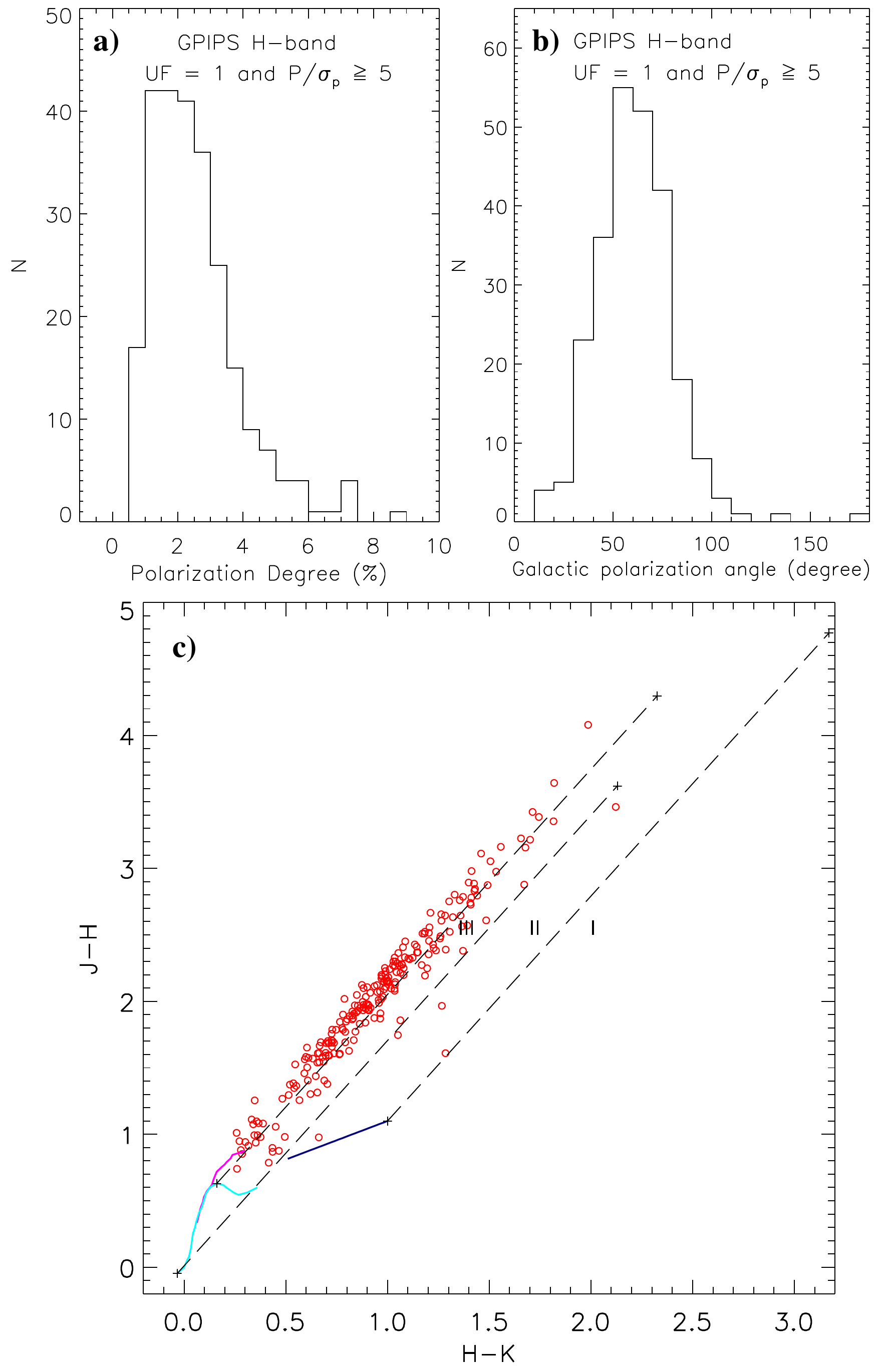}
\caption{\scriptsize a) Histogram of the GPIPS H-band polarization degree. b) The distribution of galactic position angles of the polarization vectors.
These plots are shown for stars with UF = 1 and $P/\sigma_p \ge$ 5 (also see Figure~\ref{fig8}). 
c) NIR color-color diagram (H$-$K vs J$-$H) of the selected GPIPS sources with UF = 1 and $P/\sigma_p \ge$ 5. 
The solid curves highlight the unreddened locus of main sequence stars (cyan) and giants (magenta) \citep[from][]{bessell88}. 
The three parallel long-dashed lines show the reddening vectors (with A$_{K}$ = 3 mag) of giants, main-sequence stars, 
and Classical T Tauri (CTTS) stars. 
The extinction vectors are extracted from \citet{cohen81} extinction laws (A$_{J}$/A$_{K}$ = 2.94 and 
A$_{H}$/A$_{K}$ = 1.72 for California Institute of Technology (CIT) system). 
Three different subregions (namely ``I'', ``II'', and ``III'') are marked in the diagram. 
The YSOs can be traced in ``I'' and ``II'' subregions. CTTS locus (in CIT system) \citep{meyer97} is shown 
as a solid line in navy color. The loci of unreddened dwarf (Bessell \& Brett (BB) system), 
giant (BB-system), and colors are transfered into CIT system using transformation relations given in \citet{carpenter01}. 
The plot shows that the majority of stars appears behind the N46 molecular cloud (see text for details).}
\label{fig9}
\end{figure*}

The histograms of the degree of polarization and the polarization galactic position angles of 249 stars are displayed in Figures~\ref{fig9}a and~\ref{fig9}b, respectively. The degree of polarization value for the majority of the population is found to be $\sim$3\%. Additionally, we infer an ordered plane-of-the-sky component of the magnetic field with a peak galactic position angle of $\sim$55$\degr$. In Figure~\ref{fig9}c, the GPS NIR color-color diagram of sources 
confirms that the majority of stars are located behind the N46 molecular cloud. In the NIR color-color diagram, 
the reddened background stars and/or embedded stars can be identified with (J$-$H) $\geq$ 1.0, however the foreground sources can be 
traced with (J$-$H) $<$ 1.0.
\section{Discussion}
\label{sec:disc}
\subsection{The energetics of WN7 W-R star}
\label{subsec:bub}
The presence of extended nebulous features around the O and/or W-R stars is often explained by the 
interaction between the massive stars and their surroundings \citep{deharveng10,dewangan15a,dewangan15,lamers99}. 
Many similar studies, in particular, surrounding W-R stars have been carried out in recent years 
(e.g., WR 130 \citep{cichowolski15}, WR153ab/HD 211853 \citep{vasquez10,liu12}, WR 23 \citep{cappa05}, and WR 55 \citep{cappa09}).  
The knowledge of various pressure components (pressure of an H\,{\sc ii} region $(P_{HII})$, radiation pressure (P$_{rad}$), 
and stellar wind ram pressure (P$_{wind}$)) driven by a massive star can be useful to infer the prominent physical feedback process. 
A total pressure driven by a massive star on its surroundings can be written as the sum of these three pressure components i.e. 
P$_{total}$ = P$_{H II}$ + P$_{rad}$ + P$_{wind}$. 
In recent years, it is often found that the 8 $\mu$m emission surrounds the warm dust emission as well as the 
ionized emission \citep[e.g.][]{deharveng10,paladini12}. 
In some such sites, it has been reported that the photoionized gas associated with the H\,{\sc ii} regions 
appears as the major contributor for the feedback process \citep{dewangan15a,dewangan15}.
However, in the case of bubble N46, the lack of extended ionized emission inside the bubble strongly 
indicates that the bubble is unlikely originated by the expanding H\,{\sc ii} region.
The spectroscopically characterized source WN7 W-R star appears to be located inside the bubble N46. 
Here, we estimate only two pressure components P$_{wind}$ (= $\dot{M}_{w} V_{w} / 4 \pi D_{s}^2$) and 
P$_{rad}$ (= $L_{bol}/ 4\pi c D_{s}^2$), where $\dot{M}_{w}$ is the mass-loss rate, 
V$_{w}$ is the wind velocity, L$_{bol}$ is the bolometric luminosity, and D$_{s}$ is the distance from 
the location of the W-R star where the pressure components are computed. Adopting the typical values of 
$\dot{M}_{w}$ ($\approx$ 5.0 $\times$ 10$^{-5}$ M$_{\odot}$ yr$^{-1}$) and L$_{bol}$ ($\sim$10$^{6}$ L$_{\odot}$) for WN7 W-R star 
given in \citet{lamers99} and \citet{prinja90} along with V$_{w}$ (= 1600 km s$^{-1}$) estimated 
using our observed spectra (see Section~\ref{sec:spect}), 
we obtain the values of $P_{wind}$ $\approx$ 1.1 $\times$ 10$^{-9}$ dynes cm$^{-2}$ and $P_{rad}$ $\approx$ 2.88 $\times$ 10$^{-10}$ dynes\, cm$^{-2}$ at 
a distance of 1.93 pc (i.e. an average radius of the bubble). 
We find $P_{wind}$ $>$ $P_{rad}$. Additionally, $P_{wind}$ is much higher than the pressure associated with a typical cool molecular cloud ($P_{MC}$$\sim$10$^{-11}$--10$^{-12}$ dynes cm$^{-2}$ for a temperature $\sim$20 K 
and the particle density $\sim$10$^{3}$--10$^{4}$ cm$^{-3}$) \citep[see Table 7.3 of][]{dyson80}. 
These calculations indicate that the bubble appears to be produced by the stellar wind of WN7 W-R star. 
Hence, the bubble N46 can be considered as an example of stellar-wind bubble or wind driven bubble \citep[e.g.][]{castor75}. 
From the position-velocity analysis of molecular gas, we infer the presence of an expanding shell 
with an expansion velocity, V$_{exp}$= 3 km s$^{-1}$.

We estimate the expected mechanical luminosity of the stellar wind (L$_{w}$ = 0.5$\, \dot{M}_{w}$ V$_{w}^{2}$ erg s$^{-1}$) 
for WN7 W-R star using the values of $\dot{M}_{w}$ and V$_{w}$ as mentioned above. 
We obtain L$_{w}$ $\approx$ 4 $\times$ 10$^{37}$  ergs s$^{-1}$ for WN7 W-R star.
Using the value of L$_{w}$, we infer that the W-R star has injected a large amount of mechanical energy 
(E$_{w}$ $\approx$ 1.3 $\times$ 10$^{51}$ ergs) in 1 Myr. During the entire life of a typical WN7 star (4 Myr; see Figure~13.3 given in \citet{lamers99}), 
the star can dump E$_{w}$ $\approx$ 5 $\times$ 10$^{51}$ ergs in its vicinity. 
In the environment surrounding WR153ab, \citet{vasquez10} also found that the mechanical energy released by the W-R star greatly 
influenced its vicinity, shaping the ring-like structure of the molecular distribution.  

The starlight H-band polarization data allow us to examine the large scale magnetic field structure projected in sky plane. 
The magnetic field lines are almost uniformly distributed within the N46 molecular cloud.
However, we notice a variation in the distribution of mean polarization position angles between the MME condensation and the W-R star. 
We have also found that P$_{wind}$ is the dominated term compared to P$_{rad}$. 
It is known that the aligned dust can be used as a tracer for magnetic fields. 
In the case of wind bubbles, it is often found that only some amount of the injected mechanical energy will be
transferred into kinetic energy of the expanding gas \citep{weaver77,arthur07}. 
Hence, remaining energy propagates in the surrounding interstellar medium.
Considering this fact, the impact of stellar wind of W-R star seems to be responsible for the noticeable change in the polarization angles. 
Note that, in this work, we are unable to trace the small scale magnetic field structure toward the dense regions. 
\subsection{Star formation process}
\label{subsec:signp}
In the N46 molecular cloud, star formation activities are traced by the presence of compact 
H\,{\sc ii} regions, clusters of YSOs, 6.7 GHz MME, and molecular outflow. 
In our clustering analysis, we found that the clusters of YSOs are spatially seen toward the molecular condensations, which are associated 
with a range of temperature and density of about 18--24~K and 0.6--9.2~$\times$~10$^{22}$ cm$^{-2}$ (A$_{V}$ $\sim$7--98 mag).
Some YSOs are also seen on the edges of the bubble without any clustering. 
Several compact radio peaks are identified on the edges of the bubble and are consistent with radio spectral types B0V--B0.5V.
Infrared images revealed the individual IRcs of three crss that have masses (13--20 M$_{\odot}$) and ages (1--2 Myr). 
It is evident that star formation continues into the N46 molecular cloud. 
Interestingly, our observational investigation suggests that the early phase of MSF ($<$ 0.1 Myr) is occurring 
in the MME condensation, as traced by the presence of the 6.7 GHz MME. This condensation is the most massive clump and is 
associated with significant clustering of YSOs and at least one molecular outflow. An embedded IRc of the 6.7 GHz MME is identified and is located 
within the YSOs cluster. Due to coarse beam size, the GRS $^{13}$CO data are limited and cannot allow us to study the gas kinematics within MME condensation. 
Hence, high resolution molecular line observations will be helpful to further explore the MME. We also identify an embedded 
young massive protostar (Mass $\sim$9 M$_{\odot}$, A$_{V}$ $\sim$54 mag, and age $\sim$0.7 Myr) associated with N46clm clump, which is located 
near the edge of the bubble. 
The average ages of Class~I and Class~II YSOs are generally reported to be $\sim$0.44 Myr and $\sim$1--3 Myr \citep{evans09}, respectively. 
In the N46 molecular cloud, we find the presence of evolved and powerful WN7 W-R star (with typical age $\sim$4 Myr). 
It appears that the strong stellar wind from the W-R star has affected its local vicinity, which has been inferred from the presence of a wind driven bubble/shell. 
It has been suggested that the stellar winds from O and B stars can induce the formation of massive stars in molecular clouds \citep{morton67,lucy67,lucy70}.  
We notice the existence of an apparent age gradient between the W-R star and young sources (i.e. YSOs, IRcs of crss, and 6.7 GHz MME). 
Hence, it is most likely that the stellar wind from WN7 W-R star plays a significant role in the propagation of star formation in the N46 molecular cloud. 
Previously, several authors also presented results in favour of star formation induced by the strong wind of the WR stars 
(e.g., WR 130 \citep{cichowolski15}, WR153ab/HD 211853 \citep{vasquez10,liu12}, WR 23 \citep{cappa05}, and WR 55 \citep{cappa09}).

To assess the possibility of triggered star formation, we computed $P_{wind}$ values at different distances with respect to W-R star. 
The W-R star is situated at a projected linear separation of $\sim$3.4 pc, $\sim$4.4 pc, $\sim$6.5 pc, $\sim$6.7 pc, and  $\sim$14.1 pc from 
the N46clm, clm3, clm2, MME, and clm1 molecular condensations. 
In previous section, we estimated $P_{wind}$ near the edge, 
which is equal to be $\sim$1.1 $\times$ 10$^{-9}$ dynes cm$^{-2}$ at D$_{s}$ = 1.93 pc and is much higher than $P_{MC}$. 
We also estimate $P_{wind}$ $\sim$0.94 $\times$ 10$^{-10}$ and $\sim$2.1 $\times$ 10$^{-11}$ dynes cm$^{-2}$ at D$_{s}$ = 6.7 and 14.1 pc, respectively. 
The clm1 condensation is located farthest from the W-R star, indicating the impact of wind may not be very strong. 
Considering the tremendous energy output of the W-R star, the star formation activity could be triggered on the edges of the bubble. 
Furthermore, star formation in the N46clm, clm2, clm3, and MME condensations might also be influenced by the feedback of WN7 W-R star.
However, the triggered star formation in the clm1 condensation seems unlikely. 
\section{Summary and Conclusions}
\label{sec:conc}
In this work, we have carried out a multi-wavelength analysis of a MIR bubble N46, which hosts a WN7 W-R star, 
using new observations along with publicly available archival datasets. 
The main purpose of this work is to study the star formation process in extreme 
environments (e.g., regions around OB stars and H\,{\sc ii} regions). 
The important outcomes of this multi-wavelength analysis are given below:\\
$\bullet$ Based on the FWHM of HeII 2.1885 $\mu m$ line (i.e. about 110 \AA), W-R 1503-160L is classified as a weak-lined WN7 star. 
Considering the weak-lined WN7 star and its 2MASS photometry, we estimate its distance to be 5.2$\pm$0.7 kpc, which is in good agreement with the kinematic distance of the bubble N46 (i.e. 4.8$\pm$1.4 kpc). This result leads that the bubble N46 hosts the W-R 1503-160L star.\\
$\bullet$ The molecular cloud associated with the bubble N46 (i.e N46 molecular cloud) is well traced in a velocity range of 87--100 km s$^{-1}$. 
In the integrated $^{13}$CO map, five molecular condensations (clm1, clm2, clm3, N46clm, and MME) are identified within the N46 molecular cloud.\\
$\bullet$ These molecular condensations are also traced in the {\it Herschel} maps with a range of temperature and density of about 18--24~K and 
0.6--9.2~$\times$~10$^{22}$ cm$^{-2}$ (A$_{V}$ $\sim$7--98 mag), respectively. 
The masses of the prominent clumps MME, N46clm, clm1, and clm2 are estimated to be $\sim$7273, $\sim$954, $\sim$412, and $\sim$741 $M_\odot$, respectively.\\ 
$\bullet$ The distribution of ionized emission observed in the MAGPIS 20 cm continuum map 
is mostly seen toward the edges of the bubble. 
10 crss are identified and each of them is associated with a radio spectral type B0V/B0.5V. 
Visual inspection of NIR and GLIMPSE images reveals the individual IRcs of three crss that have masses 13--20 M$_{\odot}$ and ages 1--2 Myr. \\
$\bullet$ The analysis of UKIDSS-NIR, {\it Spitzer}-IRAC and MIPSGAL photometry reveals 
a total of 370 YSOs. The surface density contours of YSOs are spatially seen toward all the molecular condensations, 
revealing ongoing star formation within the N46 molecular cloud. YSOs are also identified on the edges of the bubble without any clustering. \\ 
$\bullet$ The analysis of the gas kinematics of $^{13}$CO indicates the signature of an expanding shell with a velocity of 3 km s$^{-1}$.\\ 
$\bullet$ The MME condensation is identified as the most massive and dense clump in the N46 molecular cloud. 
The MME clump shows intense star formation activity and also contains a 6.7 GHz MME with radio continuum detection. 
An embedded IRc of the 6.7 GHz MME is identified. The MME condensation contains at least one molecular outflow. \\ 
$\bullet$ The N46clm condensation harbors an embedded protostar that has physical parameters: 
mass: 9$\pm$3 M$_{\odot}$; extinction: 54$\pm$19 mag; and age: 0.67 Myr.\\ 
$\bullet$ The pressure calculations ($P_{rad}$ and P$_{wind}$) indicate that the stellar wind associated with WN7 W-R star 
(with a mechanical luminosity of $\sim$4 $\times$ 10$^{37}$  ergs s$^{-1}$) can be considered as the major contributor for 
the feedback mechanism.\\ 
$\bullet$ A deviation of the H-band starlight mean polarization angles around the bubble also traces the feedback process. 
The strong stellar wind from WN7 W-R star (with typical age $\sim$4 Myr) has influenced its local vicinity. 
The W-R star is located at a projected linear separation of $\sim$3.4 pc, $\sim$4.4 pc , $\sim$6.5 pc , $\sim$6.7 pc, and  $\sim$14.1 pc from 
the N46clm, clm3, clm2, MME, and clm1 molecular condensations.\\ 
$\bullet$ Our analysis has revealed that star formation continues into the N46 molecular cloud. The cloud also harbors massive young stars. 
In the N46 cloud, there exists an apparent age gradient between the W-R star and young sources (YSOs, IRcs of crss, and 6.7 GHz MME).

Taking into account all the observational results obtained with our multi-wavelength analysis, 
we conclude that there is a possibility of triggered star formation toward the edges of the bubble. 
Furthermore, the star formation activities in the N46clm, clm2, clm3, and MME condensations appear to be influenced by the energetics of 
WN7 W-R star.
\newpage
\acknowledgments
We thank the anonymous reviewer for a critical reading of the manuscript and several useful comments and 
suggestions, which greatly improved the scientific contents of the paper. 
The research work at Physical Research Laboratory is funded by the Department of Space, Government of India. 
This work is based on data obtained as part of the UKIRT Infrared Deep Sky Survey. This publication 
made use of data products from the Two Micron All Sky Survey (a joint project of the University of Massachusetts and 
the Infrared Processing and Analysis Center / California Institute of Technology, funded by NASA and NSF), archival 
data obtained with the {\it Spitzer} Space Telescope (operated by the Jet Propulsion Laboratory, California Institute 
of Technology under a contract with NASA). 
This publication makes use of molecular line data from the Boston University-FCRAO Galactic
Ring Survey (GRS). The GRS is a joint project of Boston University and Five College Radio Astronomy Observatory, 
funded by the National Science Foundation (NSF) under grants AST-9800334, AST-0098562, and AST-0100793.  
This publication makes use of the Galactic Plane Infrared Polarization Survey (GPIPS). 
The GPIPS was conducted using the {\it Mimir} instrument, jointly developed at Boston University and Lowell Observatory
and supported by NASA, NSF, and the W.M. Keck Foundation. 
AL acknowledges the CONACYT(M\'{e}xico) grant CB-2012-01-1828-41. 
IZ is supported by the Russian Foundation for Basic Research (RFBR). 
\begin{deluxetable}{cccccccccccc}
\tablewidth{0pt} 
\tabletypesize{\scriptsize} 
\tablecaption{Equivalent width measurements. \label{tab2aa}}
%
\tablehead{ \colhead{Lines} & \colhead{Wavelength ($\mu$m)} & \colhead{Equivalent Width (\AA)}}
\startdata
HeII & 0.6660 &  $-$37.2\\
HeI  &0.6678 & $-$28.8 \\
HeI  &0.7065 &  $-$78.0 \\
NI   & 0.7587 &  $-$10.9 \\
NIII  &0.8236 &  $-$18.0 \\
HeII &1.6920 &  $-$39.3 \\
HeI  &1.6990 &  $-$21.5 \\
HeI  &2.0610 & $-$29.0 \\
HeI + NIII &2.1120 &  $-$49.9 \\
HeI\tablenotemark{d}  &2.1656 &  $-$58.7 \\
HeII\tablenotemark{d} &2.1890 &  $-$42.2 \\
\enddata 
 \tablenotetext{d}{These lines are blended. So their equivalent width values can be taken as indicative numbers.}
\end{deluxetable}
\begin{deluxetable}{cccccccccccc}
\tablewidth{0pt} 
\tabletypesize{\scriptsize} 
\tablecaption{Coordinates of some referred astronomical objects seen in a global view of the bubble N46, as marked in Figures~\ref{fig1} and~\ref{fig3}. \label{tab2}}
\tablehead{ \colhead{ID} & \colhead{Galactic} & \colhead{Galactic}& \colhead{RA} & \colhead{Dec}\\
\colhead{} & \colhead{Longitude (degree)}  &\colhead{Latitude (degree)}  & \colhead{[J2000]} & \colhead{[J2000]}}
\startdata
 Bubble N46 	                         &27.3100 &   $-$0.1100                      &  18:41:33.0	 &  $-$05:03:07.6\\ 								    
  W-R 1503$-$160L                   &27.2987 &   $-$0.1207                    &  18:41:34.1	  &  $-$05:04:01.2 \\ 
IRAS 18385$-$0512                  &27.1853 &   $-$0.0822                    &  18:41:13.3	 &  $-$05:09:01.1 \\ 
IRAS 18391$-$0504                  &27.3689 &   $-$0.1598                    &  18:41:50.2	  &  $-$05:01:21.0 \\ 
SNR G027.3+00.0                    &27.3866 &   $-$0.0060                    &  18:41:19.2	  &  $-$04:56:11.0 \\ 
IRDC (SDC G27.356$-$0.152)                                       &27.3568 &   $-$0.1530                    &  18:41:47.4	  &  $-$05:01:48.5 \\ 
6.7 GHz methanol maser          &27.3650 &   $-$0.1660                    &  18:41:51.0	  &  $-$05:01:42.8 \\ 
N46clm                                    &27.2940 &   $-$0.1540                    &  18:41:40.7	  &  $-$05:05:11.0 \\ 
clm1                                        &27.4653 &   $-$0.1510                    &  18:41:58.9	  &  $-$04:55:58.0 \\ 
clm2                                           &27.2418 &   $-$0.1751                    &  18:41:39.5	  &  $-$05:08:33.0 \\ 
clm3                                           &27.4653 &   $-$0.1510                    &  18:41:23.3	  &  $-$05:04:59.2  
 \enddata 
\end{deluxetable}

\begin{deluxetable}{cccccccccccc}
\tablewidth{0pt} 
\tabletypesize{\scriptsize} 
\tablecaption{Physical parameters of the crss seen on the edges of the bubble (see Figures~\ref{fig2}c). \label{tab3}}
%
\tablehead{ \colhead{ID} & \colhead{Galactic} & \colhead{Galactic}& \colhead{RA} & \colhead{Dec} & \colhead{Total flux} 	&  \colhead{logN$_{uv}$}   & \colhead{Spectral Type}\\
\colhead{} & \colhead{Longitude (degree)}  &\colhead{Latitude (degree)}  & \colhead{[J2000]} & \colhead{[J2000]}  &  \colhead{(S${_\nu}$ in Jy)} &   \colhead{(s$^{-1}$)}   &		           }
\startdata
crs1 	                         &     27.3162  &   $-$0.1349     &    18:41:39.0  &   $-$05:03:28.8       & 0.1960 		& 47.55 	 &    B0V--O9V \\ 								    
crs2                              &     27.3128  &   $-$0.1454     &    18:41:40.9  &   $-$05:03:56.9       &  0.1430	        & 47.42 	 &    B0V  \\ 
crs3                              &     27.3139  &   $-$0.1121     &    18:41:33.9  &   $-$05:02:58.4       & 0.0137	        & 46.40 	 &    B1V  \\ 
crs4                              &     27.3023  &   $-$0.1404     &    18:41:38.7  &   $-$05:04:22.4       & 0.2254	        & 47.61 	 &    B0V--O9V  \\ 
crs5                              &     27.3067  &   $-$0.1476     &    18:41:40.7  &   $-$05:04:20.1       & 0.1538	        & 47.45 	 &    B0V \\ 
crs6                              &     27.3184  &   $-$0.1232     &    18:41:36.8  &   $-$05:03:02.5       & 0.0892	        & 47.21 	 &    B0.5V--B0V  \\ 
crs7\tablenotemark{*}                              &     27.3250  &   $-$0.1204     &    18:41:36.9  &   $-$05:02:36.6       & 0.0794 		& 47.16 	 &    B0.5V--B0V \\ 
crs8                              &     27.3128  &   $-$0.1238     &    18:41:36.3  &   $-$05:03:21.2       &0.0941    		& 47.23 	 &    B0.5V--B0V  \\ 
crs9\tablenotemark{*}                              &     27.3100  &   $-$0.0949     &    18:41:29.8  &   $-$05:02:42.5       & 0.1166   		& 47.33 	 &    B0.5V--B0V  \\ 
crs10\tablenotemark{*}                           &     27.3278  &   $-$0.1132     &    18:41:35.7  &   $-$05:02:15.8       &0.0774 	       & 47.15 	        &    B0.5V--B0V  
\enddata 
 \tablenotetext{*}{The radio sources that are associated with the Infrared counterparts.}
\end{deluxetable}
\end{document}